\documentclass[acp, manuscript]{copernicus}

\renewcommand{\thesubsection}{\thesection.\alph{subsection}}

\usepackage{tikz}
\usetikzlibrary{patterns}
\usetikzlibrary{shapes}
\usetikzlibrary{decorations.text}
\usetikzlibrary{shapes.multipart}
\usetikzlibrary{arrows,shapes.geometric,positioning}
\usetikzlibrary{shapes,positioning,decorations.pathmorphing}
\usetikzlibrary{calc}
\usepackage{booktabs}
\usetikzlibrary{backgrounds}
\usepackage{varwidth}
\usetikzlibrary{fit}

\usepackage{bm,bbm}
\usepackage{amsmath,amsfonts,amssymb,mathrsfs}
\usepackage{siunitx}
\usepackage{latexsym,euscript,textcomp}
\usepackage{color,graphics,epsf,dcolumn}
\usepackage{epstopdf,ulem}
\usepackage{blindtext}
\usepackage{threeparttable}

\usepackage{enumitem}

\newcommand{\beq}{\begin{equation}}
\newcommand{\eeq}{\end{equation}}
\newcommand{\pt}{\partial}
\allowdisplaybreaks

\begin{document}
\nolinenumbers

\title{Water lifting and outflow gain of kinetic energy in tropical cyclones}


\Author[1,2]{Anastassia M.}{Makarieva}
\Author[1]{Victor G.}{Gorshkov}
\Author[1]{Andrei V.}{Nefiodov}
\Author[3]{Alexander V. }{Chikunov}
\Author[4,5]{Douglas}{Sheil}
\Author[6]{Antonio Donato}{Nobre}
\Author[7]{Paulo}{Nobre}
\Author[8]{G\"{u}nter}{Plunien}
\Author[9]{Ruben D.}{Molina}

\affil[1]{Theoretical Physics Division, Petersburg Nuclear Physics Institute, Gatchina  188300, St.~Petersburg, Russia}
\affil[2]{Institute for Advanced Study, Technical University of Munich, Lichtenbergstrasse 2 a, D-85748 Garching, Germany}
\affil[3]{Princeton Institute of Life Sciences, Princeton, New Jersey 08540, USA}
\affil[4]{Forest Ecology and Forest Management Group, Wageningen University \& Research, PO Box 47, 6700 AA, Wageningen, The Netherlands}
\affil[5]{Faculty of Environmental Sciences and Natural Resource Management, Norwegian University of Life Sciences, \AA s, Norway}
\affil[6]{Centro de Ci\^{e}ncia do Sistema Terrestre INPE, S\~{a}o Jos\'{e} dos Campos, S\~{a}o Paulo  12227-010, Brazil}
\affil[7]{Center for Weather Forecast and Climate Studies INPE, S\~{a}o Jos\'{e} dos Campos, S\~{a}o Paulo 12227-010, Brazil}
\affil[8]{Institut f\"{u}r Theoretische Physik, Technische Universit\"{a}t Dresden,  Dresden  01069, Germany}
\affil[9]{Escuela Ambiental, Facultad de Ingenier\'{i}a, Universidad de Antioquia, Medell\'{i}n, Colombia}


\runningtitle{Water lifting and outflow gain of kinetic energy in tropical cyclones}

\runningauthor{Makarieva et al.}

\correspondence{A. M. Makarieva (ammakarieva@gmail.com)}

\received{}
\pubdiscuss{} 
\revised{}
\accepted{}
\published{}


\firstpage{1}

\maketitle

\begin{abstract}
While water lifting plays a recognized role in the global atmospheric power budget, estimates for this role in tropical cyclones vary from no effect to a major reduction in storm intensity. To better assess this impact, here we consider the work output of an infinitely narrow thermodynamic cycle with two streamlines connecting the top of the boundary layer in the vicinity of maximum wind (without assuming gradient-wind balance) to an arbitrary level in the inviscid free troposphere. The reduction of a storm's maximum wind speed due to water lifting is found to decline with increasing efficiency of the cycle and is about 5\% for maximum observed Carnot efficiencies. In the steady-state cycle, there is an extra heat input associated with the warming of precipitating water. The corresponding positive extra work is of an opposite sign and several times smaller than that due to water lifting. We also estimate the gain of kinetic energy in the outflow region. Contrary to previous assessments, this term is found to be large when the outflow radius is small (comparable to the radius of maximum wind).
Using our framework, we show that Emanuel's maximum potential intensity (E-PI) corresponds to a cycle where total work equals work performed at the top of the boundary layer (net work in the free troposphere is zero). This constrains a dependence between the outflow temperature and heat input at the point of maximum wind,
but does not constrain the radial pressure gradient. We outline the implications of the established patterns for assessing real storms.
\end{abstract}

\introduction  

\label{intr}

Reliable predictions of storm intensity are vital for improving human safety. These predictions require a robust account of the major physical factors that determine the maximum wind speed that can be developed by the storm. Tropical cyclones do not just generate kinetic energy, they also lift water that subsequently precipitates. This lifting can diminish the power available for winds. Nonetheless, available estimates of this impact are inconsistent  \citep{makarieva20,er20}.

Table~\ref{EST} summarizes the situation. In steady-state large-scale circulations, the water lifting power $W_P$~(W~m$^{-2}$) is within 20-50\% of total wind power. By analogy to hydropower, this lifting power is estimated from the known precipitation rate $P$ and precipitation path length $H_P$ (the mean height from which the hydrometeors are falling) \citep{go82,go95,pa00,pa12,jas13,arxiv17}.

\begin{table}[!ht]
    \caption{Relative estimates of the contribution of water lifting to atmospheric energetics,  by different authors in chronological order. }\label{EST}
    \begin{threeparttable}
    \centering    
    \begin{tabular}{m{0.17\textwidth}m{0.14\textwidth}m{0.07\textwidth}m{0.51\textwidth}}
    \hline
     \multicolumn{1}{c}{Author} &  \multicolumn{1}{c}{Context} &  \multicolumn{1}{c}{Value ($\%$)} &  \multicolumn{1}{c}{Comment}\\  
\hline
\citet[][p.~6]{go82}, \citet[][Table~2.1]{go95} & Atmospheric circulation over land & $~~~20$ &  Precipitation over land $P = 0.5$~t~m$^{-2}$~yr$^{-1}$ \citep{lv79} falling from mid troposphere $H_P \sim 5$~km gives 
$S_L g P H_P = 1.2\times 10^{14}$~W for total land area $S_L = 1.5\times 10^{14}$~m$^2$ or $\sim 20\%$ of total atmospheric power equal to the sum of $W_P = g P H_P=0.8$~W~m$^{-2}$ and kinetic energy dissipation $W_K = 3.5$~W~m$^{-2}$. The latter value corresponds to $1\%$ contribution of total solar flux \citep[][]{gustavson79}.\\
\hline
\citet[][Fig.~3a vs. Fig.~4a]{em88}& Tropical cyclone, {\bf reversible ascent} & $~~~20$ &  A storm that lifts all condensed water from the surface with relative humidity $80\%$ and temperature $T_s = 30$\textcelsius{} to the atmospheric layer  with temperature $T_o \simeq -73$\textcelsius{} is theoretically estimated to develop a 20\% lower pressure drop ($\simeq 96$~hPa, from $1013$~hPa to $917$~hPa) than would a similar storm without lifting water ($\simeq 118$~hPa, from $1013$~hPa to $895$~hPa).\\
    \hline
\citet[][p.~1149]{em88} & Tropical cyclone, {\bf pseudo\-adiabatic ascent} & $~~~~~5$   & A storm where condensed water precipitates from the point of condensation is theoretically estimated to develop a 5\% lower pressure drop than would a similar storm without lifting water.\\
\hline
\citet{pa00,pa12} & Tropical convection & $~~~50$ & Satellite-derived $W_P = 1.5$~W~m$^{-2}$  \citep{pa12} and $W_K = 1.4$~W~m$^{-2}$ \citep{pa00} correspond to  $W_P/(W_P + W_K) \times 100\% \simeq 50\%.$ \\
\hline
\citet[][Fig.~4a]{sabuwala15} & Tropical cyclone, {\bf pseudo\-adiabatic ascent} & $~~~50$ & A hurricane with maximum velocity $\simeq 50$~m~s$^{-1}$ is theoretically estimated to develop a $50\%$ lower squared maximum velocity than would a similar storm without lifting water. \\ 
\hline
\citet{jas13,arxiv17} & Global atmospheric circulation & $~~~30$ &  $W_K = 2.3$~W~m$^{-2}$ based on instantaneous MERRA re-analysis data and $W_P = 1$~W~m$^{-2}$ based on observations and theoretical estimates correspond to $W_P/(W_P + W_K) \times 100\% \simeq 30\%.$ \\
\hline
\citet{re19} & Tropical cyclone, {\bf reversible cycle} & $~~~~~0$ & Condensed water does not precipitate but travels with the air parcels both up and down, net water lifting power \citep[the integral over $dq_t$ in Eq.~(13) of][]{re19} is zero.\\
\hline
This work & E-PI cyclone with $\varepsilon_C \simeq 0.3$ & $~~~10$ & Cumulative account of water lifting and water warming, Eq.~(\ref{K1}), reduces squared maximum 
velocity by about $10\%$ in both reversible and pseudoadiabatic storms.\\
\hline
    \end{tabular}
\begin{tablenotes}[para,flushleft]
\end{tablenotes}
\end{threeparttable}
  \end{table}

For tropical cyclones, \citet{em88} estimated that water lifting  reduces the central pressure drop in intense storms by about $5\%$ and $20\%$ for pseudoadiabatic and reversible ascent, respectively, and concluded that {\textquotedblleft}the importance of  water loading in limiting the hurricane intensity in the reversible case{\textquotedblright} is {\textquotedblleft}very substantial{\textquotedblright}. Without referring to this prior work, \citet{er20} recently agreed with \citet{makarieva20} that in real cyclones the reduction of the squared maximum velocity due to water lifting should not  exceed $10\%$.

In contrast, \citet{sabuwala15} --quoted by \citet{em18} but neglected by \citet{re19} and by \citet{er20}-- used satellite-derived precipitation data and Emanuel's potential intensity  framework to report an approximately 50\% reduction
in the squared maximum velocity due to water lifting for pseudoadiabatic ascent. Unlike \citet{sabuwala15}, who did not quote \citet{em88}, \citet{wang20} used the approach of \citet{em88} to account for the total water mixing ratio $q_t$ in the pseudoadiabatic model of \citet{emanuel11} and found that this reduces air velocity at the radius of maximum wind in a hurricane with reversible adiabats by about $10\%$  (or squared velocity by $20\%$). At the same time, \citet{er20} indicated that the impact of the water lifting  on storm intensity depends on the integral of $dq_t/dt$ over a closed contour.  For a reversible cycle, which conserves the total water content, this integral is exact zero (Table~\ref{EST}).

The preceding issues raise several questions. First, is the water lifting impact on storm intensity large or small, and if it is small, why is this different from the power budget of larger-scale circulations? Second, what is the reason for the high observation-derived estimates of \citet{sabuwala15}? Third, why is the impact of the water lifting maximized in reversible compared with pseudoadiabatic hurricanes?

Assessing the influence of water lifting on a storm's steady-state intensity 
requires a consideration of the storm's thermodynamic cycle. The original derivation of a storm's maximum velocity by \citet{em86} was based on a scaling relation between velocity and temperature along a surface of constant moist saturated entropy and angular momentum  (\citealt[][Eq.~(13)]{em86}; \citealt[][Eq.~(11)]{emanuel11}). The derivation assumed the  free troposphere to be in gradient-wind balance. \citet[][their Fig.~1, Eqs.~(p5) and (p6)]{makarieva18} showed that this assumption can be relaxed  in the assessment of storm-integrated energy fluxes. Kerry Emanuel suggested\footnote{K. Emanuel made this suggestion in his signed review of (subsequently rejected) submission of \citet{makarieva18} to the Journal of Geophysical Research: Atmospheres.} that  \citet{makarieva18}'s approach could be used locally to describe an infinitely narrow cycle in the vicinity of maximum wind. Without referring to \citet{makarieva18}, \citet{re19} applied this suggestion but, as noted by \citet{ms20} and \citet{makarieva20}, their derivations were based on an incorrect configuration of air streamlines.

Here we consider an infinitely narrow thermodynamic cycle in the vicinity of maximum wind  that comprises two streamlines connecting the top of the boundary layer to some arbitrary level in the free troposphere  (Fig.~\ref{fig1}). Our analysis assumes the atmosphere to be inviscid above the boundary layer, but does not require the gradient-wind balance 
at the radius of maximum wind. We show that the expression for work of this cycle is equivalent to the scaling relation in the original \citet{em86}'s derivation (Section~\ref{frame}). The new, more general formulation of E-PI framework is useful in the following three aspects.

First, by allowing an explicit evaluation of the water lifting term that we perform in Section~\ref{ests}, it responds to the above three questions concerning the contribution of water lifting to storm's energetics.

Second, it allows the estimation of the gain of kinetic energy and angular momentum in the outflow region of the storm. Regarding this term, it has long been held that it can be large only when the outflow radius is very large (\citealt[][p.~602]{em86}; \citealt[][p.~190]{emanuel04}).  \citet{re19} noted that the outflow term  {\textquotedblleft}will be small if the radius at which this occurs is not too large{\textquotedblright}. However, \citet{makarieva18b}, see also \citet{makarieva20}, showed that, conversely,  this term {\textquotedblleft}is significant when the outflow radius ... is close to the radius of maximum wind{\textquotedblright}, i.e., when the outflow radius is small. Omitting to quote \citet{makarieva18b} or to discuss their own previous opposing view, \citet{er20} made an effort to re-derive the result of \citet{makarieva18b} about the (in)significance of the outflow term at (large) small outflow radii. \citet{er20}'s derivations were not conclusive, however, as they based on their Eq.~(6), where the dimensions of the right-hand and left-hand sides do not match.  As \citet{makarieva20} argued, this is due to an incorrect transition from volume to surface power fluxes. Here a consistent derivation of the outflow term is presented (Section~\ref{ests}).

Third, the new formulation demonstrates that E-PI at the point of maximum wind corresponds to a thermodynamic cycle with zero work in the free troposphere (Section~\ref{kk}). This strong constraint, together with the recently revealed relation between the inner core and outflow parameters in E-PI, is essential for evaluating "superintensity" (hurricane wind speeds exceeding their E-PI limits) \citep{mn21}.

\section{An infinitely narrow thermodynamic cycle}
\label{frame}

\subsection{Combining dynamics and thermodynamics}
\label{cdt}

\begin{figure*}[!ht]
\begin{minipage}[p]{1\textwidth}
\centering\includegraphics[width=0.5\textwidth,angle=0,clip]{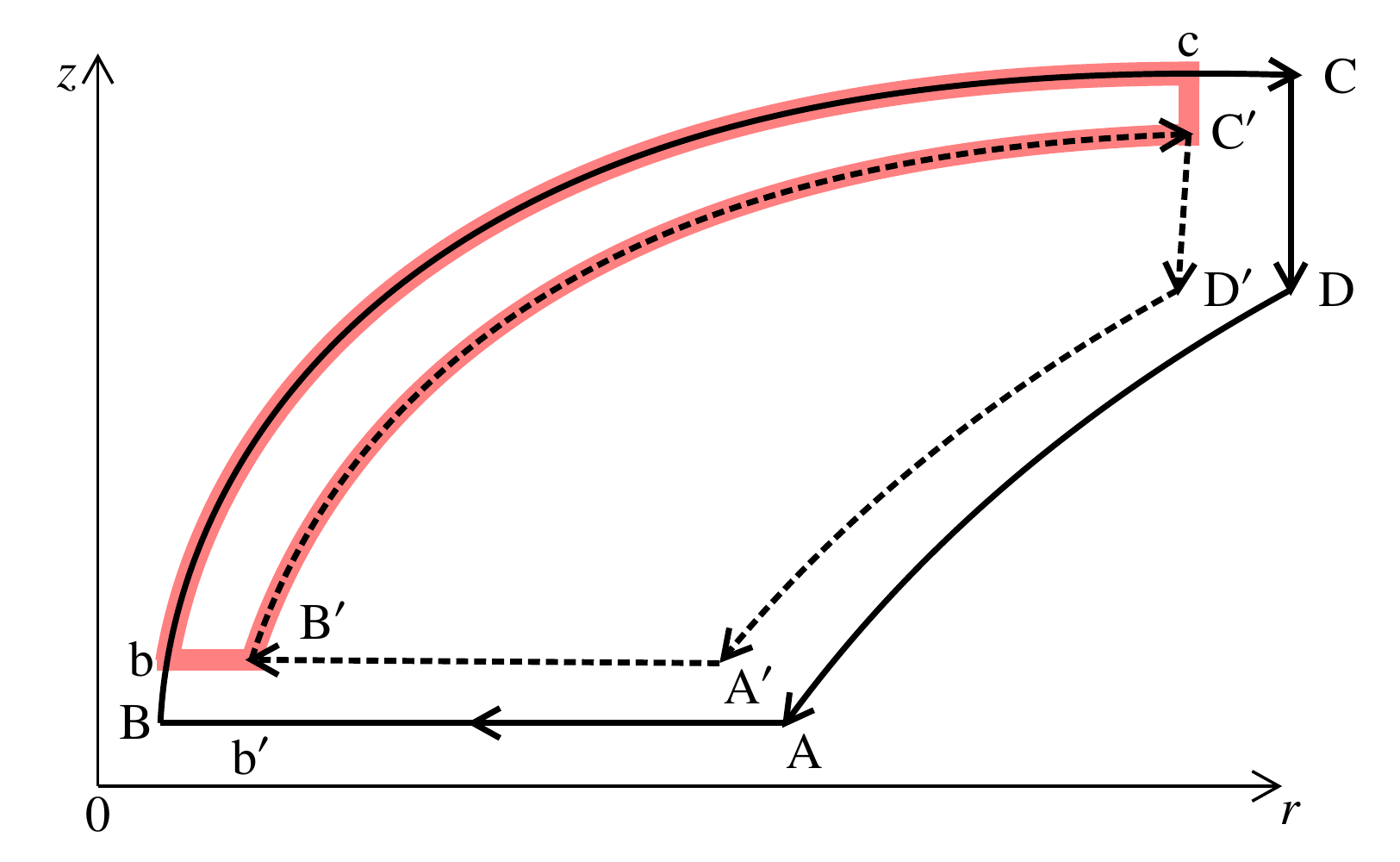}
\end{minipage}
\caption{
Air streamlines (solid and dotted black arrows) and the infinitely narrow thermodynamic cycle (thick pink lines) considered in the text \citep[cf. Fig.~1b of][]{makarieva20}. The $z$ and $r$ axes correspond to the altitude above the sea level and the distance from hurricane center, respectively;  $r_{\rm b'} = r_{\rm B'}$, $r_{\rm c} = r_{\rm C'}$.  Points $\rm B$ and $\rm B^{\prime}$ are infinitely close and chosen in the vicinity of maximum wind. The atmosphere is inviscid for $z > z_{\rm b} = z_{\rm B'}$.}
\label{fig1}
\end{figure*}

We consider two closed air streamlines, $\rm ABCDA$ and $\rm A'B'C'D'A'$ (Fig.~\ref{fig1}), in an axisymmetric atmosphere.  The goal of our derivations is to find the relation between turbulent dissipation and heat input at the top of the boundary layer. From this relation, the maximum wind speed in E-PI can be estimated. 

We assume hydrostatic equilibrium and apply the Bernoulli equation to path $\rm b'Bb$ that belongs to streamline $\rm ABCDA$:
\begin{gather}\label{he}
-\alpha \frac{\partial p}{\partial z} = g,\\
\label{B}
-\alpha dp = d\left(\frac{V^2}{2}\right) +g dz - \mathbf{F} \cdot d\mathbf{l},
\end{gather}
where $\alpha \equiv 1/\rho$, $\rho = \rho_d + \rho_v + \rho_l$ is the density of moist air (including dry air $\rho_d$, water vapor $\rho_v$ and condensed water $\rho_l$), $p$ is air pressure, $g$ is the acceleration of gravity, $V$ is air velocity,  $\mathbf{F}$ is the turbulent friction force per unit air mass, and $d\mathbf{l}=\mathbf{V}dt$.

The connection between the dynamics and thermodynamics will be found through the common term $\alpha dp$. The logic of our derivations is schematized in Fig.~\ref{fig2}.

\tikzset{
state/.style={draw,rectangle split,rectangle split parts=3,
              rectangle split part fill={red!10,blue!7,blue!7},text width=10cm, text centered,minimum height=2cm}
}

\tikzset{
statet/.style={draw, rectangle split,rectangle split parts=2,
              rectangle split part fill={red!10,blue!5},text centered,minimum height=2.1cm}
}

\tikzset{
statet4/.style={draw,rectangle split, rectangle split parts=4,
              rectangle split part fill={red!10,blue!5,red!10,blue!5},text width=10cm, text centered,minimum height=2.1cm}
}

\tikzset{
boxx/.style={draw, rectangle,fill={red!10}, text centered,minimum height=2.1cm}
}

\newdimen\nodcv
\nodcv=0.5cm
\newdimen\nodch
\nodch=0.07cm
\newdimen\mh
\mh=6cm
\newdimen\fp
\fp=3.4cm
\newdimen\tw
\tw=7.2cm
\newdimen\bh
\bh=0.85cm

\tikzset{pblock/.style = {rectangle split, minimum height=1.8cm, rectangle split horizontal,
                      rectangle split parts=2,  rectangle split part fill={red!10,blue!4},very thick,draw=lightgray, align=center,execute at begin node=\setlength{\baselineskip}{12pt}}}

\tikzset{pblock1/.style = {rectangle split, minimum height=1.8cm, rectangle split horizontal,
                      rectangle split parts=2,  rectangle split part fill={green!10,pink!5},very thick,draw=lightgray, align=center,execute at begin node=\setlength{\baselineskip}{12pt}}}

\begin{figure}[ht!]

\begin{tikzpicture}[node distance = \nodch and \nodcv,font=\small]
\linespread{0.66}

\node at (0,-1) (oo){};

\node [pblock1, left= of oo](Fdl) {\nodepart[text width=\fp]{one}Lower atmosphere:\\Bernoulli equation,\\hydrostatic equilibrium,\\$V_{\rm b'} = V_{\rm B'}$ \\
\nodepart[text width=\tw+1.4cm]{two}
\vspace{-0.3cm}
\begin{gather}\label{Fdl}
-\int_{\rm b'}^{\rm b} {\mathbf F} \cdot d{\mathbf l} = -\int_{\rm B'}^{\rm b} \alpha dp -\int_{\rm B'}^{\rm b} d\left(\frac{V^2}{2}\right)
\end{gather}
};

\node [below=2.2\bh of Fdl.east, node distance = 0cm](bFdl){};

\node [pblock1, left=1.5\nodch of bFdl.east](adp) {\nodepart[text width=\fp]{one}Upper atmosphere:\\Bernoulli equation,\\hydrostatic equilibrium\\ at ${\rm cC'}$, ${\mathbf F} = 0$ \\
\nodepart[text width=\tw]{two}
\vspace{-0.3cm}
\begin{gather}\label{adp}
-\int_{\rm B'}^{\rm b} \alpha dp  = -\oint \alpha dp + \int_{\rm B'}^{\rm b} d\left(\frac{V^2}{2}\right) + \int_{\rm c}^{\rm C'} d\left(\frac{V^2}{2}\right)
\end{gather}
};

\node [below=2.2\bh of adp.east, node distance = 0cm](badp){};

\node [pblock1, left=1.5\nodch of badp.east](q) {\nodepart[text width=\fp]{one}Wet and dry mass units:\\ $(1+q_t)\alpha \equiv \alpha_d$,\\$(1+q_t)\delta Q  \equiv \delta Q_d$\\
\nodepart[text width=\tw-1.5cm]{two}
\vspace{-0.3cm}
\begin{gather}\label{qt}
 -\oint \alpha dp \equiv  -\oint \alpha_d dp + \oint q_t \alpha dp
\end{gather}
};

\node [below=2.2\bh of q.east, node distance=0 cm](bq){};

\node [pblock1, left=1.5\nodch of bq.east](eps) {\nodepart[text width=\fp]{one}Thermodynamics:\\ Link between work and heat input\\
\nodepart[text width=\tw-3cm]{two}
\vspace{-0.3cm}
\begin{gather}\label{dQ}
-\oint \alpha_d dp = \varepsilon \int_{\rm B'}^{\rm b} \delta Q_d
\end{gather}
};

\node [below=9.1\bh of Fdl.west](beps){};

\node[pblock,right=0cm of beps.south](rev){\nodepart[text width=\fp+1.4cm]{one} Link between\\ volume-specific\\frictional dissipation\\ and heat input\\   \nodepart[text width=\tw]{two}\vspace{-0.3cm}
\begin{gather}\label{rel}
-\int_{\rm b'}^{\rm b} {\mathbf F} \cdot d{\mathbf l} = \varepsilon \int_{\rm B'}^{\rm b} \delta Q_d +  \oint q_t \alpha dp + \int_{\rm c}^{\rm C'} d\left(\frac{V^2}{2}\right)
\end{gather}};

\node [below=11.4\bh of Fdl.west](feps){};

\node[pblock,right=0cm of feps.south](fin){\nodepart[text width=\fp+1.4cm]{one} Reversible cycle with $q_t = \rm const$,\\ no kinetic energy change \\ at $\rm B'b$ (point of maximum wind)  \\ and in 
the outflow  $\rm cC'$\\ \nodepart[text width=\tw]{two}\vspace{-0.3cm}
\begin{gather}\label{relm}
-\int_{\rm B'}^{\rm b} \alpha dp =  -\oint \alpha dp = \varepsilon \int_{\rm B'}^{\rm b} \delta Q 
\end{gather}};

\end{tikzpicture}
\vspace{-0.2cm}
\caption{Key steps of deriving the relation between turbulent dissipation and heat input in the lower atmosphere. The integrals over a closed contour refer to $\rm B'bcC'B'$ in Fig.~\ref{fig1}. The heat input $\delta Q_d$ is normalized per unit dry air mass.
}
\label{fig2}
\end{figure}


Applying Eqs.~(\ref{he}) and (\ref{B}) to $\rm b'Bb$   results in Eq.~(\ref{Fdl}) (Fig.~\ref{fig2}), which relates work of the friction force to the sum of the  horizontal differences in pressure and kinetic energy  per unit moist air mass (halved squared velocity). In Eq.~(\ref{Fdl}), we have additionally assumed that $V_{\rm b'} = V_{\rm B'}$. This implies two possibilities.  One is that $\pt V/\pt z = 0$, which holds by definition at the point of maximum wind and is otherwise a plausible assumption at the top of the boundary layer, where turbulent viscosity becomes negligible  \citep[][p.~3045]{bryan09b}. Another possibility is that points $\rm b'$ and $\rm B'$ (and, respectively, $\rm B$ and $\rm b$) coincide, such that path $\rm b'b$ is horizontal. This second case with $z_{\rm b} = 0$ was considered by \citet[][their Fig.~1]{re19}, who assumed that $|\mathbf{F}| = 0$ for $z > 0$ but $|\mathbf{F}| \ne 0$ at $\rm B'b$. For a derivation of Eq.~(\ref{Fdl}) from the equations of motion, see \ref{A0}.

Our next step is to consider the inviscid atmosphere above $z = z_{\rm b}$. Applying the Bernoulli equation (\ref{B}) with $|\mathbf{F}|=0$ to streamlines $\rm bc$ and $\rm B'C'$ and assuming hydrostatic equilibrium at path $\rm cC'$ (which is not a streamline) yields Eq.~(\ref{adp}). It relates the horizontal change of $\alpha dp$ at $\rm B' b$ to the sum of the integral of $\alpha dp$ over the closed contour ${\rm B'bcC'B'}$ and the changes of kinetic energy  at $\rm B' b$ and in the outflow region $\rm cC'$ (Figs.~\ref{fig1} and \ref{fig2}). (When these changes are zero, the first equality of Eq.~(\ref{relm}) follows.)

At this point, we invoke the relation between the specific volumes of moist and dry air, $\alpha_d = (1+q_t)\alpha$, where $\alpha_d \equiv 1/\rho_d$ is the specific volume of dry air, and $q_t \equiv (\rho_v + \rho_l)/\rho_d$ is the total water mixing ratio. This relation allows us to link the integrals of $\alpha dp$ and $\alpha_d dp$ over the closed contour $\rm B'bcC'B'$, Eq.~(\ref{qt}).

On the other hand, the integral of $\alpha_d dp$ over a closed contour represents work done per unit dry air mass
in the corresponding thermodynamic cycle. This work is converted from the heat input with the cycle's efficiency $\varepsilon$, Eq.~(\ref{dQ}).  By summing Eqs.~(\ref{Fdl})--(\ref{dQ}) we combine the dynamic and thermodynamic constraints to obtain a relation between turbulent dissipation and heat input in the lower atmosphere, Eq.~(\ref{rel}). 

In the particular case of $q_t= \rm  const$, one can divide the functions under the integral signs in Eq.~\eqref{dQ} by a constant factor $1+q_t$. Then  we obtain the second equality in Eq.~\eqref{relm}. 

In the derivation, the hydrostatic equilibrium approximation (\ref{he}) was applied at $\rm b'Bb$ and $\rm cC'$, but it was not used at $\rm bc$ and $\rm B'C'$. 
The thermodynamic processes in the cycle were not specified, so Eq.~(\ref{dQ}) can be viewed as defining the value of $\varepsilon$.
The choice of the outflow point $\rm c$ along the streamline in Eqs.~\eqref{adp} and \eqref{rel} was arbitrary. The assumption that the cycle is infinitely narrow was used in  Eqs.~\eqref{Fdl} and \eqref{rel}, but not in Eqs.~\eqref{adp}--\eqref{dQ}.
Equation \eqref{relm} describes an infinitely narrow cycle $\rm B'bcC'B'$ with $\pt V^2/\pt z = 0$ at $\rm cC'$ and $\pt V^2/\pt r = 0$ at $\rm B'b$.  It is also valid for a special
case of $\rm B'bcC'B'$ being a {\it closed streamline} with $V_{\rm b} = V_{\rm B'}$ \citep{em88}.

\subsection{Conventional E-PI estimate}
\label{em}

We will now demonstrate the equivalence of the framework depicted in Fig.~\ref{fig2} to E-PI in two ways: in terms  of turbulent dissipation and in terms of angular momentum.

The ratio of the surface fluxes of turbulent dissipation and ocean-to-atmosphere heat is proportional to squared velocity \citep[e.g.,][Eqs.~(15) and (16)]{bister98}. An independent estimate of this ratio would yield a constraint on velocity. Such an estimate can be deduced from Eq.~(\ref{rel}) with some assumptions.

\citet{em86} did not discriminate between $\alpha$ and $\alpha_d$ (and accordingly between $\delta Q$ and $\delta Q_d$) and thus neglected the second term on the right-hand side of Eq.~(\ref{rel}).   The third term on the right-hand side of Eq.~(\ref{rel}), which is the change of kinetic energy in the outflow, was also neglected.  That was because \citet{em86} assumed gradient-wind balance and, hence, $V = v$, where $v$ is tangential velocity, and then chose point $\rm c$ in the outflow where $V = v = 0$ and, hence, $\pt V^2/\pt z = 0$.
Finally, \citet{em86} considered the thermodynamic cycle to be reversible, such that its efficiency $\varepsilon$ equals Carnot efficiency  
$\varepsilon_C \equiv (T_{\rm b}-T_{\rm c})/T_{\rm b}$.

Applying these assumptions --neglecting the last two terms and putting $\varepsilon = \varepsilon_C$ and $\delta Q_d = T_{\rm b} ds^*$ in Eq.~(\ref{rel})--
we lift the integral signs in the limit of the infinitely narrow cycle $\rm b' \to B'$ and divide both sides of the equation by $dt$, to obtain:
\beq\label{FVe1}
-\mathbf{F}\cdot\mathbf{V} = \varepsilon_C \frac{\delta Q_d}{dt} = \varepsilon_C T_{\rm b} \frac{\pt s^*}{\pt r}u,
\eeq
where $\mathbf{V} = d\mathbf{l}/dt$ is total air velocity, $u = dr/dt$ is radial velocity, and $s^*$ is moist saturated entropy.

Multiplied by $\rho$, Eq.~(\ref{FVe1}) relates local volume-specific rates  (W~m$^{-3})$ of turbulent dissipation and  heat input into a horizontally expanding air parcel. Assuming that the ratio of these volume-specific rates is the same as the ratio of the corresponding surface fluxes (W~m$^{-2})$ of turbulent dissipation $D = \rho C_D V^3$ and heat input $J = \rho C_k V (k_s^* - k)$, 
\beq\label{as1}
- \frac{\mathbf{F}\cdot\mathbf{V}}{\delta Q_d/dt} = \frac{D}{J},
\eeq
yields the original E-PI formula for maximum velocity \citep[e.g.,][Eq.~(22)]{emanuel11}:
\beq\label{vmax}
V_{\rm max}^2 =\frac{D}{J}\frac{C_k}{C_D} (k_s^* - k)= \varepsilon_C \frac{C_k}{C_D}(k_s^* - k) .
\eeq
Here $C_k \simeq C_D$ are surface exchange coefficients for enthalpy and momentum, respectively; $k_s^*$ (J~kg$^{-1}$) is the saturated enthalpy of air at surface temperature, and $k$ is the actual enthalpy of air.

The second way of demonstrating the equivalence between E-PI and the framework in Fig.~\ref{fig2} is to show that
the following combination of Eqs.~(\ref{adp})--(\ref{dQ}),
\beq
\label{eq}
-\int_{\rm B'}^{\rm b} \alpha dp -\int_{\rm B'}^{\rm b} d\left(\frac{V^2}{2}\right) = \varepsilon \int_{\rm B'}^{\rm b} \delta Q_d +  \oint q_t \alpha dp + \int_{\rm c}^{\rm C'} d\left(\frac{V^2}{2}\right),
\eeq
is, under E-PI assumptions, equivalent to \citet{emanuel11}'s Eq.~(11),
\beq\label{E11}
\frac{\nu_1 - \nu_2}{T_1 - T_2}= -\frac{ds^*}{dM}, \quad  (z \ge z_{\rm b}),
\eeq
where $T_1$, $T_2$  and $\nu_1 \equiv v_1/r_1$, $\nu_2 \equiv v_2/r_2$ are, respectively, air temperatures and angular velocities at arbitrary distances $r_1$ and $r_2$ from the storm center on a surface of constant angular momentum $M$ and 
moist saturated entropy $s^* = s^*(M)$ defined by the given value of $ds^*/dM$. 

Using the definition of angular momentum 
\begin{equation}\label{M}
M \equiv vr + \frac{f r^2}{2},
\end{equation}
where the Coriolis parameter $f \equiv 2 \Omega \sin \varphi$ is assumed constant ($\varphi$ is latitude, $\Omega$ is the angular velocity of Earth's rotation), and assuming gradient-wind balance at point $\rm b$,
\begin{equation}\label{gwb}
\alpha \frac{\partial p}{\partial r} = \frac{v^2}{r} + fv,
\end{equation}
we can write our Eq.~(\ref{eq}) (with the second term on the right-hand side ignored) as
\beq\label{eq2}
\int^{\rm B'}_{\rm b} \frac{v}{r}\frac{\pt M}{\pt r}  dr = -\varepsilon \int^{\rm B'}_{\rm b} \delta Q_d +  \frac{1}{2}(v^2_{\rm C'}-v^2_{\rm  c}).
\eeq
Here we have assumed, as did \citet{er20}, that the change of velocity $V$ over path $\rm cC'$ is dominated by the change in tangential velocity $v$, 
$V^2_{\rm C'}-V^2_{\rm  c} = v^2_{\rm C'}-v^2_{\rm  c}$. Since the atmosphere at $\rm cC'$ is frictionless and hydrostatic, 
it is equivalent to assuming local gradient-wind balance (see Eq.~\eqref{A10}). With path $\rm cC'$ hydrostatic
and in gradient-wind balance, all our results are invariant with respect to its orientation (whether/how $\rm cC'$ is tilted about the vertical axis).

To lift the integral signs and describe an infinitely narrow cycle, we need to evaluate the last term in Eq.~(\ref{eq2}) in the limit $r_{\rm B'} \to r_{\rm b}$. Since the atmosphere is frictionless above $z_{\rm b}$ and since the pressure field is axisymmetric, paths $\rm bc$ and $\rm B'C'$ conserve angular momentum $M$.  Using Eq.~(\ref{M}) for $M$ in the equation $M_{\rm B'} - M_{\rm b} = M_{\rm C'} - M_{\rm c}$ and dividing this equation by  $r_{\rm c} (r_{\rm B'} - r_{\rm b})$ we obtain
\begin{equation}\label{Vo4}
\frac{1}{r_{\rm c}}\left[ v_{\rm B'} + r_{\rm b} \frac{v_{\rm B'} - v_{\rm b}}{r_{\rm B'} - r_{\rm b}}  + \frac{f}{2} (r_{\rm B'} + r_{\rm b}) \right] = \frac{v_{\rm C'} - v_{\rm c}}{r_{\rm B'} - r_{\rm b}}.
\end{equation}
Taking the limit $r_{\rm B'} \to r_{\rm b}$ and multiplying Eq.~(\ref{Vo4}) by $v_{\rm c}$ we find
\begin{equation}\label{Vo5}
v_{\rm c} \frac {\partial v_{\rm c}}{\partial r} = \frac{v_{\rm c}}{r_{\rm c}} \left( v + r \frac{\partial v}{\partial r} + fr \right)  = \frac{v_{\rm c}}{r_{\rm c}}\frac{\pt M}{\pt r}, \quad (z = z_{\rm b}).
\end{equation}

We now assume that our infinitely narrow cycle has Carnot efficiency $\varepsilon_C = (T_{\rm b} - T_{\rm c})/T_{\rm b}$
and that the heat input at $\rm B'b$ can be expressed in terms of the increment of moist saturated entropy $\delta Q_d = T_{\rm b} ds^*$.  Lifting the integral signs in Eq.~(\ref{eq2}) with the use of Eq.~(\ref{Vo5}), we obtain
\beq\label{eq3}
 \frac{v_{\rm b}}{r_{\rm b}} \frac{\pt M}{\pt r}  = - (T_{\rm b} - T_{\rm c}) \frac{\pt s^*}{\pt r} +\frac{v_{\rm c}}{r_{\rm c}}\frac{\pt M}{\pt r}, \quad (z = z_{\rm b}).
\eeq
Assuming, finally, that $s^* = s^*(M)$, such that $ds^*/dM = (\pt s^*/\pt r)/(\pt M/\pt r)$, we obtain Eq.~(\ref{E11}) from Eq.~(\ref{eq3}) with points $1$ and $2$ corresponding to points $\rm b$ and $\rm c$, respectively.

With point $\rm c$ chosen such that $v_{\rm c} = 0$, Eqs.~(\ref{E11}) and (\ref{eq3}) are equivalent to \citet{em86}'s Eq.~(13)  \citep[see also][Eq.~(12)]{emanuel11}. 
\citet{em86} made several assumptions about the boundary layer
(specifically, that the surfaces of constant $s^*$ and $M$  are
vertical and that the horizontal turbulent diffusion fluxes are small)
to justify that the radial gradients of $s^*$ and $M$  relate as their
surface fluxes $\tau_s= J/T_s$ and $\tau_M= - D r /V$, respectively:
\beq\label{as2}
\frac{\pt s^*/\pt r}{\pt M/\pt r} = \frac{\tau_s}{\tau_M}.
\eeq
Then assuming that $V \simeq v_{\rm b}$,  $r \simeq r_{\rm b}$, and $T_s \simeq  T_{\rm b}$, where $T_s$ is sea surface temperature, Eqs.~(\ref{eq3}) 
and (\ref{as2}) yield Eq.~(\ref{vmax}). Under Eq.~(\ref{eq3}), assumptions (\ref{as1}) and (\ref{as2}) are equivalent.

Equation~(\ref{vmax}) relates local fluxes and is intended to estimate storm's maximum potential intensity.  However, neither Eq.~(\ref{FVe1}) nor Eq.~(\ref{eq3}), from which the maximum potential intensity (\ref{vmax}) can be derived, require $\partial V^2/\partial r = 0$ (the condition of maximum wind). This peculiarity of E-PI was noted by \citet[][]{montgomery17b}.  Equation~(\ref{FVe1}) requires $\pt V^2/\pt z = 0$ at $z = z_b$, while Eq.~(\ref{eq3}) does not.

As we will discuss in Section~\ref{kk}, $\pt V^2/\pt r = 0$ is an important constraint on E-PI. Here we note that, according to Eq.~(\ref{relm}), in the E-PI framework work in the free troposphere (along the path ${\rm bcC'B'}$) is zero. The total work of the cycle, i.e., heat input at $\rm B'b$ multiplied by efficiency $\varepsilon_C$,  equals the work on $\rm B'b$. Since $\varepsilon_C$ depends on the outflow temperature $T_{\rm c}$,  the specification of the thermodynamic process at $\rm B'b$ and the choice of $T_{\rm c}$ cannot be independent \citep{mn21}.

\section{Estimating the water lifting and outflow terms}
\label{ests}

\subsection{Reversible and pseudoadiabatic hurricanes}
\label{revh}

For the considered thermodynamic cycle $\rm B'bcC'B'$ to have Carnot efficiency, it should be reversible.
This requires that the air is saturated (relative humidity $\mathcal{H}$ is equal to $100\%$) and the total moisture mixing ratio $q_t$ is constant, everywhere in the cycle\footnote{In the literature one can sometimes find a loose definition of a reversible process that only assumes $q_t = \rm const$ but allows the relative humidity to vary \citep[e.g.,][Eq.~(23)]{bryan09a}.}.

With $q_t = \rm const$, dividing both sides of Eq.~(\ref{dQ}) by  $(1 + q_t)$ we obtain
\beq\label{adpm}
-\oint \alpha dp = \varepsilon \int_{\rm B'}^{\rm b} \delta Q,
\eeq
where $\delta Q \equiv \delta Q_d/(1+q_t)$ is the heat input per unit moist air mass. Summing Eqs.~(\ref{Fdl}), (\ref{adp}) and (\ref{adpm}) yields
\beq\label{rev}
-\int_{\rm b'}^{\rm b} {\mathbf F} \cdot d{\mathbf l} = \varepsilon \int_{\rm B'}^{\rm b} \delta Q +  \int_{\rm c}^{\rm C'} d\left(\frac{V^2}{2}\right).
\eeq
Comparison of Eqs.~(\ref{rel}) and (\ref{rev}) shows that the latter lacks the water lifting term. \citet{re19} similarly found that the water lifting term  is comprised in the integral of the material derivative $dq_t/dt$  over a closed streamline, which is zero when $q_t = \rm const$ (Table~\ref{EST}). The physical meaning of this result is that all the water that is lifted in the ascending branch of the cycle taking the energy away, goes down in the descending branch and performs work, with the net effect being zero.

However, if the storm circulation is composed of streamlines representing reversible processes  (saturated isotherms and adiabats conserving $q_t$), but $q_t$ differs between streamlines, the thermodynamic cycle $\rm B'bcC'B'$ will {\it not} be reversible due to the change of $q_t$ on paths $\rm B'b$ and $\rm cC'$ that connect different streamlines,  ${\rm ABCDA}$ and ${\rm A'B'C'D'A'}$. The efficiency of such a cycle will be lower than Carnot efficiency.

\subsection{Extra heat input to warm precipitating water}
\label{cl}

A cycle with reversible adiabats $\rm bc$ and $\rm C'B'$ ($q_t = \rm const$, $\mathcal{H} = 100\%$)
and saturated isotherms $\rm B'b$  and $\rm cC'$, along which $q_t$, respectively, increases and decreases, 
has the following relation between work and heat input (for derivations, see \ref{eh}):
\beq\label{stc}
-\oint \alpha_d dp \simeq \varepsilon_C T_{\rm b} \Delta s^* + \frac{\varepsilon_C}{2} c_l (T_{\rm b} - T_{\rm c})\Delta q_t,
\eeq
where $\varepsilon_C \equiv (T_{\rm b} - T_{\rm c})/T_{\rm b}$ is Carnot efficiency, $\Delta q_t \ge 0$ and $\Delta s^* \ge 0$ are, respectively, the changes of total  water mixing ratio $q_t$ and moist saturated entropy $s^*$ from $\rm B'$ to $\rm b$, $c_l = 4.2$~kJ~kg$^{-1}$~K$^{-1}$ is the specific heat capacity of liquid water.  The first term  on the right-hand side of  Eq.~\eqref{stc} is due to a heat input into an expanding air parcel with evaporating water.  If, as we assume below, all water at the warmer isotherm is added in the form of water vapor, then $\Delta q_t$ should be replaced by saturated water vapor mixing ratio $\Delta q^*$ in Eq.~(\ref{stc}).

While in finite differences Eq.~(\ref{stc}) is valid only when $\rm B'b$ and $\rm cC'$ are isotherms, in the limit of an infinitely narrow cycle $\rm B'bcC'B'$, when $\rm B'b$ and $\rm cC'$ degenerate each to a point, Eq.~(\ref{stc}) becomes valid even if the temperature along $\rm B'b$ and $\rm cC'$ is not constant (see \ref{eh}). 
As one of our reviewers pointed out, this is due to the small change of temperature along these paths compared to the finite difference $T_{\rm b} - T_{\rm c}$
\citep[see also][p.~59]{carnot1890}.
This explains how \citet{em86} obtained a Carnot efficiency multiplier at the radial heat input $T_{\rm b} \pt s^*/\pt r$ in his Eq.~(13)  without assuming horizontal isothermy at
the top of the boundary layer.

The last term in Eq.~(\ref{stc}), see also Eq.~(\ref{oints1}), corresponds to term {\textquotedblleft}(c){\textquotedblright} in Eq.~(19) of \citet{em88}, who described it  as {\textquotedblleft}the increase of entropy due to addition of water mass{\textquotedblright} and {\textquotedblleft}the contribution of water substance to the heat capacity{\textquotedblright}. Without referring to \citet{em88}, \citet{pa11} re-derived this term in his Eq.~(B2) and interpreted it as {\textquotedblleft}additional work{\textquotedblright}\footnote{Equation~(B3) of \citet{pa11} should have $T_{\rm in}$ instead of $T_{\rm out}$ in the denominator of the right-hand part, otherwise it contradicts Eq.~(B2) from which supposedly derives.}, accounting for which elevates the cycle's efficiency above Carnot efficiency.  \citet{pa11} explained that this elevation does not violate the second law of thermodynamics because the cycle is open (moisture is added and removed), while Carnot efficiency limits the efficiencies of closed cycles only. On the other hand, according to \citet{pa11}, it is not accidental that the same cycle has Carnot efficiency when $c_l = 0$: it is because this open cycle is thermodynamically equivalent to a closed cycle where the moisture removed at the colder isotherm is kept within the heat engine and added back to the cycle at the warmer isotherm. The question to this interpretation is why with $c_l \ne 0$ such a cycle is not equivalent to a closed one.

This is resolved by recognizing an additional heat input to the cycle. Warming the water removed at the colder isotherm with temperature $T_{\rm c}$ and returned at the warmer isotherm with $T_{\rm b}$ requires extra heat $c_l (T_{\rm b} - T_{\rm c})\Delta q^*$. 
As the moist air ascends and cools from $T_{\rm b}$ to $T_{\rm c}$, the mean temperature at which the water loses heat is, in the linear approximation, 
 $T_{\rm w} \simeq (T_{\rm b} + T_{\rm c})/2$.
Accordingly, the maximum efficiency with which this heat can be converted to work is $(T_{\rm b} - T_{\rm w})/T_{\rm b} \simeq \varepsilon_C/2$.

Thus, such a cycle is equivalent to a closed cycle with total heat input
\beq\label{dQd}
\Delta Q_d = T_{\rm b} \Delta s^* + c_l (T_{\rm b} - T_{\rm c}) \Delta q^*
\eeq	
and efficiency
\beq\label{eps}
\varepsilon = \varepsilon_C \left[1 - \frac{1}{2} \frac{c_l (T_{\rm b} - T_{\rm c}) \Delta q^*}{T_{\rm b} \Delta s^* + c_l (T_{\rm b} - T_{\rm c}) \Delta q^*}\right] , 
\eeq
that is lower than Carnot efficiency ($\varepsilon_C/2 \leq \varepsilon \leq\varepsilon_C$) due to the irreversibility associated with warming the precipitating water. This inherent  thermodynamic imperfection of steam cycles was recognized already by Sadi~\citet[][p.~58]{carnot1890}.

In a pseudoadiabatic hurricane, all condensed water is immediately removed (precipitates) from the air parcel: $\rho_l = 0$ and $q_t = q^*$. The extra heat reduces to $c_l (T_{\rm b} - T_P)\Delta q^*$, where $T_P$ is the mean temperature at which condensation and precipitation occur ($T_{\rm c} \leq T_P < T_{\rm b}$).  The last term in Eq.~(\ref{stc}) can be roughly  approximated by $c_l (T_{\rm b} - T_P)^2\Delta q^* /(2T_{\rm b})$.

\subsection{Water lifting}
\label{wlft}

Taking into account that paths $\rm B'b$ and $\rm cC'$ are, respectively, horizontal and vertical (Fig.~\ref{fig1}) and using $\alpha_d = (1+q^*) \alpha$ for $\rm B'b$, we can combine Eqs.~(\ref{Fdl}), (\ref{dQ}), (\ref{rel}) and (\ref{stc})
in the following form:
\begin{equation}\label{est0}
\int_{\rm b}^{\rm B'} \left( \alpha \frac{\partial p}{\partial r} + \frac{1}{2}\frac{\partial V^2}{\partial r} \right)dr  
	= -\varepsilon_C \int_{\rm b}^{\rm B'} T\frac{\partial s^*}{\partial r} dr - c_l T_{\rm b}\frac{\varepsilon^2_P}{2}  \int_{\rm b}^{\rm B'} \frac{\partial q^*}{\partial r}dr  +\oint \alpha q_t dp + \int_{\rm c}^{\rm C'} d \left( \frac{V^2}{2}\right),
\end{equation} 
where $\varepsilon_P \equiv 1 - T_P/T_{\rm b}$. Here we have assumed that evaporation occurs from the ocean surface (there is no condensate along $\rm B'b$: $q_t = q^*$). We calculate the water lifting term in \ref{wl} 
and lift the integral signs in Eq.~(\ref{est0}) for the infinitely narrow cycle $\rm B'bcC'B'$, i.e., considering the limit $r_{\rm B'} \to r_{\rm b}$ and $z_{\rm C'} \to z_{\rm c}$. This gives for $z = z_{\rm b}$ and $r = r_{\rm b}$
\begin{equation}\label{alg}
(1+q^*) \left( \alpha \frac{\partial p}{\partial r} + \frac{1}{2}\frac{\partial V^2}{\partial r} \right) =
-\varepsilon_C T_{\rm b}\frac{\partial s^*}{\partial r} 
+ \left[\frac{1}{2}\left( V_P^2 -  V^2\right) +g (z_P - z_{\rm b}) -c_l T_{\rm b}\frac{\varepsilon^2_P}{2}  \right]\frac{\pt q^*}{\pt r}  + (1+q_{t \rm c})\frac{1}{2}\frac{\partial V_{\rm c}^2}{\partial r}.
\end{equation}
Here  $q_{t \rm c}$ and $V_{\rm c}$ refer to the point $(r_{\rm c}, z_{\rm c})$;  $\pt V_{\rm
c}^2/\pt r$ is the limit of $(V^2_{\rm C'} - V^2_{\rm c})/(r_{\rm B'}
- r_{\rm b})$ at  $r_{\rm B'} \to r_{\rm b}$ and $z_{\rm C'} \to
z_{\rm c}$, cf. Eqs.~(\ref{Vo4}) and (\ref{Vo5}). The quantities
$V^2_P$,  $z_P$  and  $T_P$ are evaluated on the unclosed contour $\rm
bcC'B'$ (denoted by $\curvearrowright$):
\beq\label{def1}
V_P^2 \equiv - \frac{1}{\Delta q^*} \int_{\curvearrowright} V^2 dq_t,
\quad z_P \equiv - \frac{1}{\Delta q^*} \int_{\curvearrowright} z
dq_t,
\quad T_P \equiv - \frac{1}{\Delta q^*} \int_{\curvearrowright} T dq_t,
\eeq
where $\Delta q^* \equiv q^*_{\rm b} - q^*_{\rm B'}$. In a cycle with reversible adiabats,  i.e., with constant $q_t$ along $\rm bc$ and $\rm C'B'$ and $\Delta q^*= q_{t\rm c} - q_{t\rm C'}$, the integrals in Eq.~(\ref{def1}) 
in the considered limit are straightforwardly evaluated: $V_P= V_{\rm c}$, $z_P= z_{\rm c}$, and $T_P= T_{\rm c}$.

Equation  (\ref{alg}) summarizes the energy budget of the infinitely narrow cycle $\rm B'bcC'B'$ and  thus provides a relation between local variables. The first and second terms in the square brackets represent kinetic and potential energy increments associated with phase transitions. Term $(V_P^2-V^2)/2$ formally accounts for water vapor being added to the air mixture with kinetic energy $V^2/2$ of the air at $\rm B'b$, while disappearing via precipitation with kinetic energy $V_P^2/2$ in the free troposphere. We will neglect this term\footnote{Term $(V_P^2-V^2)/2$ can be explicitly accounted for by specifying the interaction between condensate and air (i.e., introducing a specific term to the equations of motion and Bernoulli equation).  If condensate is assumed to have the same horizontal velocity as air \citep[see, e.g.,][]{oo01,jgra17}, then as the precipitating condensate leaves the air at the surface, it will have the same velocity as the newly evaporated water vapor. The net contribution of this term to the total power budget of the hurricane will be exact zero \citep[see][Fig.~1]{arxiv17}.}, since for typical $z_P \sim 10$~km and $V \sim 60$~m~s$^{-1}$, $V^2/2$ is only about 2\% of $gz_P$.

The water lifting term $g(z_P-z_{\rm b})$ accounts for the net energy expended to lift water. 
In a real cycle, where there is no precipitation in the descending branch $\rm C'B'$ and $q_t$ does not change, $z_P$ equals the mean precipitation height $H_P$
in the ascending branch $\rm bc$. In an infinitely narrow cycle with reversible adiabats,  $q_t$ is
also constant in the "descending branch", and $z_P = z_{\rm c} = H_P$ has the same meaning.

However, in an infinitely narrow cycle with $q_t$ varying along $\rm bc$ and $\rm C'B'$, 
moisture disappears along $\rm bc$. It then arises anew along $\rm C'B'$
with its own non-zero gravitational energy.
In this hypothetical cycle, moisture performs work as it descends along $\rm C'B'$ and consumes energy when it is raised along $\rm bc$. 
Thus, the net energy $g(z_P-z_{\rm b}) \Delta q^*$ is equal to the difference in the gravitational energy of "precipitation" between $\rm bc$ and $\rm C'B'$
\citep[cf.][Eq.~(C12)]{em88}.

This net energy is equal to the energy $g (H_P - z_{\rm b}) \Delta
q^*$ spent to raise the {\it additional} moisture $\Delta q^*$ from
$z_{\rm b}$ to the mean precipitation height $H_P$ along $\rm bc$ plus
the energy $g \Delta q^*   \pt H_P/ \pt q^{*}_{\rm b}  (q^*_{\rm b} -
q^*_{\rm c})$ spent to raise {\it nearly all moisture} $q^*_{\rm b} -
q^*_{\rm c} \sim q^*_{\rm b}$  to an {\it additional} altitude $\Delta
q^*   \pt H_P/ \pt q^{*}_{\rm b}$ equal to the difference between the
mean precipitation heights along $\rm bc$ and $\rm B'C'$. For
pseudoadiabats, $H_P$ is approximately proportional to $q^*_{\rm b}$,
so that  $q^*_{\rm b} \pt H_P/ \pt q^*_{\rm b} \sim H_P$. In this
case, $z_P-z_{\rm b}$ turns out to be approximately twice the mean
precipitation height, i.e., around $10$~km for $T_{\rm b} \simeq
300$~K, Eq.~\eqref{Acalc}.

The third term in the square brackets represents the warming of the precipitating water, Eq.~(\ref{dQd}). 
It is of the opposite sign to the water lifting term. For reversible adiabats $T_P = T_{\rm c}$; 
with $T_{\rm b} = 300$~K, $T_{\rm b} - T_P \simeq 100$~K and $z_P - z_{\rm b} \simeq 17$~km (which corresponds
to mean lapse rate $\Gamma = (T_{\rm b} - T_P)/(H_P - z_{\rm b}) \simeq 6$~K~km$^{-1}$), 
the water warming term constitutes about 40\% of the water lifting term.
Accounting for water warming somewhat, but not fully, compensates the impact of water lifting, and the more so, 
the larger the difference $T_{\rm b} - T_P$. 
For pseudoadiabats, $T_P$ is equal to the mean temperature of precipitating water (Eq.~\ref{Acalc}),
the difference $T_{\rm b} - T_P$ is relatively small, and the impact of water warming is almost negligible.
With $T_{\rm b} = 300$~K, $T_{\rm b} - T_P = 25$~K and $z_P - z_{\rm b} = 10$~km (Table~\ref{revnum}), it does not exceed $10\%$.

In units of latent heat of vaporization $L_v$, the cumulative contribution $K_1$
of the water lifting and warming 
\beq\label{K1}
K_1 \equiv  \frac{1}{L_v}\left[g(z_P-z_{\rm b})  - c_l T_{\rm b}\frac{\varepsilon_P^2}{2} \right] 
\eeq
is of similar magnitude $K_1 = 0.04$ for reversible and pseudoadiabatic storms with high $\varepsilon_C \geq 0.3$ (Table~\ref{revnum}).

\subsection{The outflow gain of kinetic energy}
\label{out}

The last term on the right-hand side of Eq.~(\ref{alg}) describes how the kinetic energy in the outflow, $V_{\rm c}^2/2$, changes depending on the radius $r=r_{\rm b}$ where the streamline crosses the level $z = z_{\rm b}$ (Fig.~\ref{fig1}).  
Assuming that we are at a point near the radius of maximum wind, such that $v/r \gg \pt v/\pt r$, we can neglect $\pt v/\pt r$ in Eq.~(\ref{Vo5}) and use
the relation between tangential velocity and pressure gradient
\beq\label{Bsup}
\alpha\frac{\pt p}{\pt r} \equiv \mathcal{B} \left(\frac{v^2}{r} + fv\right),
\eeq
where $\mathcal{B}>0$ defines the degree to which the air flow is radially unbalanced ($\mathcal{B} < 1$ for the supergradient flow when the outward-directed centrifugal force 
exceeds the inward-pulling pressure gradient),
to obtain from Eq.~(\ref{Vo5}):
\beq\label{vc1}
\frac{1}{2}\frac{\partial V_{\rm c}^2}{\partial r}\simeq 
\frac{1}{2}\frac{\partial v_{\rm c}^2}{\partial r} = \frac{v_{\rm c}}{r_{\rm c}}\frac{r}{v} \left( \frac{v^2}{r} + fv \right) = \alpha\frac{\pt p}{\pt r} \frac{1}{\mathcal{B}}\frac{v_{\rm c}}{r_{\rm c}}\frac{r_{\rm b}}{v_{\rm b}},  \quad  ( r = r_{\rm b},   z = z_{\rm b}).
\eeq

On the other hand, using Eqs.~(\ref{M}) and the fact that $M_{\rm b} = M_{\rm c}$ we can express $v_{\rm c}$ as
\begin{equation}\label{voo}
v_{\rm c} = \frac{1}{r_{\rm c}} \left[v_{\rm b}r_{\rm b} + \frac{f}{2} (r_{\rm b}^2 - r_{\rm c}^2)\right].
\end{equation}
(Evaluating the derivative of Eq.~(\ref{voo}) with respect to $r_{\rm b}$ and taking into account that $\partial r_{\rm c} /\partial r_{\rm b} = 0$ 
is another way to obtain Eq.~(\ref{Vo5})). Using Eq.~(\ref{voo}) and neglecting $q_{t\rm c} \ll 1$ we can write the last term in Eq.~(\ref{alg}) as $K_2 \alpha \pt p/\pt r$,
where 
\beq\label{K2}
K_2 \equiv \frac{1}{\mathcal{B}} \frac{v_{\rm c}}{r_{\rm c}}\frac{r_{\rm b}}{v_{\rm b}} =
\frac{1}{\mathcal{B}} \left[
\frac{r_{\rm b}^2}{r_{\rm c}^2} + \frac{fr_{\rm b}}{2v_{\rm b}}\left(\frac{r_{\rm b}^2}{r_{\rm c}^2}-1\right)\right].
\eeq
For $(r_{\rm b}/r_{\rm c})^2 \ll 1$, this term is small and negative. In contrast, \citet[][p.~602]{em86} incorrectly concluded that the
outflow term becomes significant if the outflow radius is very
large\footnote{This conclusion stemmed from
\citet{em86}{\textquoteright}s Eq.~(18), where an outflow term
proportional to a large squared radius first appeared. While deriving
this equation for a cycle with finite $\rm B'b$,
\citet[][p.~588]{em86}, on the one hand, used the conservation of
angular momentum along streamlines $\rm bc$ and $\rm B'C'$ and, on the
other hand, assumed $r_{\rm B'}$ (interpreted as
{\textquotedblleft}the radial extent of the storm near the sea
level{\textquotedblright}) to be large enough for $\pt p/\pt r$ to
vanish and, at the same time, small enough for  $r \pt p/\pt r |_{\rm
b} \gg  r \pt p/\pt r |_{\rm B'}$\citep[for details,
see][Appendix~A]{makarieva18b}. With $r \pt p/ \pt r \sim \rho v^2$,
ignoring this term at $\rm B'$ means that $V^2_{\rm b} - V^2_{\rm B'}
\sim V^2_{\rm b}$, while at the point of maximum wind this difference
is zero.}.
In their re-evaluation of this issue, \citet{er20} considered  the material derivative of angular momentum $dM/dt$ along the path $\rm cC'$ connecting the two streamlines. 
Their derivation is not valid, since $\rm cC'$ is not a streamline and the air does not move along that path. Defending their configuration of streamlines, \citet{er20}  noted that {\textquotedblleft}the properties of  $\rm D'$ and $\rm D$, and of $\rm A'$ and $\rm A$ are identical{\textquotedblright} but said nothing about the properties of $\rm C'$ and $\rm C$ \citep[see Fig.~1 of][]{re19}.

\subsection{Estimates of maximum velocity}
\label{vmaxn}

We will now consider the point where $\pt V/\pt r = 0$ (see Eq.~(\ref{relm}) in Fig.~\ref{fig2}). This corresponds to the point of maximum wind if point $\rm b$ is chosen at the top of the boundary layer as in the derivations of \citet{em86} and \citet{emanuel11} or to the point  of maximum surface wind if point $\rm b$ is chosen at the surface as in the derivations of \citet{re19}. 

By analogy with Eq.~(\ref{FVe1}), from Eq.~(\ref{Fdl}) with $\pt V/\pt r = 0$ we obtain \citep[see also][Eq.~(14)]{makarieva20}:
\beq\label{FV}
-\mathbf{F}\cdot\mathbf{V} = -\alpha \frac{\pt p}{\pt r} u,
\eeq
where $u=dr/dt$. Equation~(\ref{FV}) is fundamental: it derives from the Bernoulli equation and hydrostatic equilibrium (Fig.~\ref{fig2}). It predicts that if at the point of maximum wind $|\mathbf{F}| \to 0$ (as is the case at the top of the boundary layer), then $u \to 0$. This feature is observed in numerical models \citep[e.g.,][p.~3054]{bryan09b}. It is also valid for a horizontal streamline at $z_{\rm b} = 0$ as considered by \citet[][their Fig.~1]{re19}, whereby $u \ne 0$ and $|\mathbf F| \ne 0$.

On the other hand, for $\pt v/\pt r = 0$ we have from Eqs.~(\ref{M}) and (\ref{Bsup})
\beq\label{Bsup2}
\alpha \frac{\partial p}{\partial r} = \mathcal{B} \frac{v}{r} \frac{\pt M}{\pt r}.
\eeq

Finally, using the definitions of $K_1$ (\ref{K1}) and $K_2$ (\ref{K2}) and neglecting $q^* \ll 1$, we can write
Eq.~(\ref{alg}) for $\pt V/\pt r = 0$ as follows:
\beq\label{algn}
-\alpha \frac{\partial p}{\partial r} = 
\frac{\varepsilon_C  - \beta K_1}{1 - K_2} T_{\rm b} \frac{\pt s^*}{\pt r}, \quad \beta \equiv \frac{L_v \pt q^*/\pt r}{T_{\rm b} \pt s^*/\pt r}.
\eeq
Here $0 \le \beta \le 1$ is the share of latent heat in total heat input into the air parcel along $\rm B'b$.

Now using two distinct assumptions, (\ref{as1}) and (\ref{as2}), about how the volume and surface energy fluxes relate,
we obtain two expressions for $V_{\rm max}$ that differ by a factor of $\mathcal B$.
From Eq.~(\ref{algn}), Eq.~(\ref{Bsup2}) and Eq.~(\ref{as2}) we obtain
\beq\label{Vmaxn}
V_{\rm max}^2 = \frac{\varepsilon_C  - \beta K_1}{\mathcal{B}(1 - K_2)}\frac{C_k}{C_D}(k_s^* - k). 
\eeq
From Eq.~(\ref{algn}) multiplied by $u$, Eq.~(\ref{FV}) and Eq.~(\ref{as1}) we obtain the same expression, 
but without $\mathcal{B}$. Since $\mathcal{B}$ can be as small as $0.5$ \citep[e.g.,][Fig.~8]{bryan09b}, this is a significant source
of uncertainty in $V_{\rm max}$ associated with assumptions (\ref{as1}), (\ref{as2}) and their modifications as discussed elsewhere \citep{mn21}. 

For our present purpose of estimating the role of the water lifting and the outflow, this does not matter,
since the values of $K_1$ and $K_2$ are compared with $\varepsilon_C$ and unity, respectively.
With $K_1 \simeq 0.04$, $\beta \sim 0.7$ estimated from a typical Bowen ratio, and $\varepsilon_C = 0.3$,
$K_1$ reduces $V_{\rm max}^2$ by $\beta K_1/\varepsilon_C \times 100\% \simeq 10$\% and $V_{\rm max}$ by $5\%$
both for pseudoadiabatic and adiabatic cases at $T_{\rm b} = 300$~K.

This result is comparable to \citet{em88},  who found that the central pressure drop is reduced by water lifting by about $20\%$  in the reversible case (Table~\ref{EST}). \citet{em88} did not evaluate local maximum velocity but considered a large-scale thermodynamic cycle with a horizontally isothermal top of the boundary layer. Horizontal isothermy is generally not compatible with the other E-PI assumptions and leads to an underestimate of $\alpha \pt p/\pt r$ \citep{mn21}.
This could cause the overestimate of $K_1$ for the reversible case. The details of pseudoadiabatic calculations were not reported by \citet{em88}; 
we can hypothesize that neglecting the second term in Eq.~(\ref{Acalc}) for $z_P$ could cause the underestimate of $K_1$ for the pseudoadiabatic case.

As for the outflow term $K_2$, since $r_{\rm b} < r_{\rm c}$, the factor in square brackets in Eq.~(\ref{K2}) is confined between $-f r_{\rm b}/(2 v_{\rm b})<0$ and unity.  
For characteristic values of $r_{\rm b} = 30$~km, $v_{\rm b} = 60$~m~s$^{-1}$, $\varphi = 15^{\rm o}$  and $f \simeq 3.77 \times 10^{-5}$~s$^{-1}$ we have $f r_{\rm b}/(2v_{\rm b}) \simeq  10^{-2}$. With $\mathcal{B} \simeq 1$, $K_2$ reduces $V_{\rm max}^2$ by about $1\%$ for large $r_{\rm c}$.

On the other hand, if the outflow radius is relatively small, $K_2$ is positive and {\it elevates} rather than lowers the maximum velocity estimate.  This may happen for many storms with $r_{\rm b}/r_{\rm c} \gtrsim \sqrt{f r_{\rm b}/(2v_{\rm b})}\sim  10^{-1}$.
Interpreting $K_2$ as {\textquotedblleft}dissipation{\textquotedblright} 
to occur at an arbitrary point $\rm c$ in an otherwise frictionless troposphere, is incorrect \citep[cf.][]{er20}. When $K_2 > 0$ the kinetic energy increases from $\rm c$ to $\rm C'$.  \citet{smith14} noted that a mechanism that would provide an increment of angular momentum, and hence a kinetic energy increment, along $\rm B'b$ does not appear to exist. But \citet{ar17} indicated that extra angular momentum can arise in the upper atmosphere as a real steady-state hurricane is an open system that moves through the atmosphere and can import angular momentum as it imports air and water vapor. 

\citet[][their Eq.~(2), the {\textquotedblleft}adiabatic case{\textquotedblright}]{sabuwala15} did not derive their formulations from 
the original assumptions of E-PI. However, they included what seemed a plausible water lifting term in the hurricane's power budget, as follows:
\beq\label{DS0}
\varepsilon_C (D + J) = D + g H_P P,
\eeq
from which one obtains
that
\beq\label{DS1}
V_{\rm max}^2 = \frac{D}{J}\frac{C_k}{C_D} (k_s^* - k) =\frac{1}{1 - \varepsilon_C}\left(\varepsilon_C-\frac{PL_v}{J}\frac{gH_P}{L_v}\right)\frac{C_k}{C_D} (k_s^* - k).
\eeq
\noindent
Here $P$ (kg~m$^{-2}$~s$^{-1}$) is the local precipitation in the region of maximum winds\footnote{Note the following differences in notations between \citet{sabuwala15} and the present work:  $T_s \to T_{\rm b}$, $\dot{Q}_{in} \to J$, $\dot{Q}_{d}\to D$, $P \to W_P = g H_P P$. Factor $1/(1-\varepsilon_C)$ is due to dissipative heating.}. Equation~(\ref{DS1}) is obtained  from \citet{sabuwala15}'s Eq.~(2) and their additional equation $\dot{Q}_{in} - \dot{Q}_{out} = P$. (The {\textquotedblleft}diabatic case{\textquotedblright} of \citet{sabuwala15} addressed thermal dissipation of the potential energy of the falling droplets  \citep[see][]{igel18}.) 

Compared to (\ref{Vmaxn}), the water lifting term $gH_P/L_v \sim K_1$ in Eq.~(\ref{DS1}) is multiplied by a large factor $PL_v/J \gg 1$ reflecting the ratio of local precipitation to local heat input. For typical Bowen ratios in hurricanes $B \equiv J_S/J_L \simeq  1/3$ \citep[e.g.,][]{jaimes15} we have $J = (1 + B)J_L = (1+B) EL_v$ and $PL_v/J = (P/E)/(1+B)$, where $J_S$, $J_L$ and $E$ are the local fluxes of sensible heat, latent heat, and evaporation, respectively.
Ratio $P/E$ between local precipitaton and evaporation in the region of maximum winds is variable but on
average of the order of $10$ \citep[see][their Table~1 and Figs.~2 and 3]{ar17}. 
For $H_P \sim 5$~km and $\varepsilon_C \sim 0.3$, \citet{sabuwala15}'s correction to $V_{\rm max}^2$ is thus $10 (gH_P/L_v)/[(1+B) \varepsilon_C] \simeq 0.5$, 
which is five times larger than our $\beta K_1/\varepsilon_C \simeq 0.1$ (Table~\ref{EST}).
The unjustified replacement of evaporation by precipitation caused \citet{sabuwala15}'s estimate to be too high.

\section{The physical meaning of E-PI at the point of maximum wind}
\label{kk}

Equation~(\ref{FV}) shows that, at the point of maximum wind, the local volume-specific rate (W~m$^{-3}$) of turbulent dissipation, $-\rho \mathbf{F}\cdot\mathbf{V}$,  is equal to the local volume-specific rate of sensible heat input  into an isothermally and horizontally expanding air parcel, $-(\pt p/\pt r)u$. Thus, if all this turbulent dissipative power transforms locally to heat, the external sensible heat input into the air parcel must be zero.

In the general, non-isothermal case, we can relate turbulent dissipation to latent heat input. Inspecting our key equation (\ref{alg}), we notice that the water lifting term in Eq.~(\ref{alg}) is proportional to $\pt q^*/\pt r$, while
the outflow term is proportional to $\partial p/\partial r$, see Eq.~(\ref{vc1}). On the other hand,  using the definition of $q^*$ and Clausius-Clapeyron law, the radial gradient of moist saturated entropy in Eq.~(\ref{alg}) can also be expressed  in terms of $\pt q^*/\pt r$ and $\partial p/\partial r$, see Eqs.~(\ref{Tds}) and (\ref{dT}):
\begin{equation}
\label{dsq}
T \frac{\pt s^*}{\pt r} = L_v  \frac{\pt q^*}{\pt r} \left(1+ \varkappa_1 \right) -\alpha_d \frac{\pt p}{\pt r} \left(1 - \varkappa_2 \right).
\end{equation}
Coefficients $\varkappa_1$ and $\varkappa_2$ describe the deviation from horizontal isothermy: 
\begin{gather} \label{tau1}
\varkappa_1 \equiv  \frac{1}{\mu \gamma^*_d \xi^2 (1 + \gamma^*_d)} = \frac{\varkappa_2}{\gamma^*_d \xi} \simeq  0.3, \\ \label{tau2}
\varkappa_2 \equiv \frac{1}{\mu \xi (1+ \gamma^*_d)} \simeq \frac{1}{\mu \xi} \simeq 0.2,
\end{gather}
where $\xi \equiv L/RT \simeq 18$, $\mu \equiv R/(c_p M_d) \simeq 2/7$ and $\gamma^*_d \equiv (M_d/M_v) q^*$ are defined in \ref{eh}.  The numerical values correspond to $T = 300$~K and $\gamma^*_d \equiv p_v^*/p_d = 0.04$. For $\pt T/\pt r = 0$, the terms proportional to $\varkappa_1$ and  $\varkappa_2$ {\it jointly} vanish from Eq.~(\ref{dsq}), see also Eq.~(\ref{dT}). Therefore, the isothermal case can be formally obtained  by putting $\varkappa_1 =0$ and $\varkappa_2 = 0$ in Eq.~(\ref{dsq}) and related.

Using Eqs.~(\ref{vc1}) and (\ref{dsq}) we write Eq.~(\ref{alg}) with $\pt V^2/\pt r = 0$ and $q^* \ll 1$ neglected, as follows
\begin{equation}\label{alg1}
\alpha \frac{\partial p}{\partial r} =
\left[ -\varepsilon_C  \left( 1+ \varkappa_1 \right)
+K_1 \right] L_v\frac{\pt q^*}{\pt r}  + \left[ \varepsilon_C \left(1  - \varkappa_2\right) +K_2 \right] \alpha \frac{\partial p}{\partial r}.
\end{equation}
Multiplying Eq.~(\ref{alg1}) by $-u$ and using Eq.~(\ref{FV}), we obtain
\beq\label{sum}
-\mathbf{F}\cdot\mathbf{V}  = \frac{\varepsilon_C (1+ \varkappa_1) - K_1}{1 - \varepsilon_C (1 - \varkappa_2) - K_2} L_v \frac{\pt q^*}{\pt r} u.  
\eeq
(This equation together with Eq.~(\ref{algn}) show that E-PI constrains the latent-to-total heat ratio $\beta$ at $\rm B'b$.
For $K_1 = K_2 = \varkappa_1 = \varkappa_2 = 0$ we have $\beta = 1 - \varepsilon_C$, which, as can be verified from the exact relation (\ref{beta}), 
is a good approximation of $\beta$ for small $K_1 \ll \varepsilon_C$.) Assuming that turbulent dissipation relates to {\it latent} heat input
equally at $\rm bB'$ and at the surface (i.e., replacing in
Eq.~(\ref{as1}) total heat $T ds^*$ and $J$ with latent heat $L_vdq^*$
and $J_L =  \rho C_k V L_v (q^*_s - q)$, respectively),
\beq\label{as3}
-\frac{\mathbf{F}\cdot\mathbf{V}}{L_v (\pt q^*/\pt r) u} = \frac{D}{J_L},
\eeq
from Eq.~(\ref{sum}) we obtain another expression for $V^2_{\rm max}$:
\beq\label{Vmaxn2}
V_{\rm max}^2 = \frac{D}{J_L} \frac{C_k}{C_D} L_v(q_s^* - q) =
\frac{\varepsilon_C (1+ \varkappa_1) - K_1}{1 - \varepsilon_C (1 -
\varkappa_2) - K_2} \frac{C_k}{C_D}L_v (q_s^* - q).
\eeq
When surface sensible heat $J_S = \rho C_k V c_p (T_s - T)$ is
negligibly small, such that $J_L \gg J_S$ and $J = J_L+J_S \simeq J_L$
\citep[as assumed by][who put $T = T_s$]{em86},
Eq.~(\ref{Vmaxn2}) coincides with the {\textquotedblleft}dissipative
heating{\textquotedblright} formulation
\citep[][Eq.~(21)]{bister98}. 
Indeed, for $K_1 = K_2 = \varkappa_1 = \varkappa_2 = 0$, Eq.~(\ref{Vmaxn2}),
derived without assuming any dissipative heating,
is then equivalent to Eq.~(\ref{vmax}) but with $\varepsilon_C$ replaced
in the latter by $\varepsilon_C/(1 - \varepsilon_C)$. 
This formal similarity encountered in a numerical model
could lead to the dissipative heating formulation.

While Eq.~(\ref{sum}) relates turbulent dissipation to latent heat input alone, this does not mean that {\textquotedblleft}only surface latent heat fluxes can power tropical cyclones{\textquotedblright}, which is how \citet{er20} apparently misunderstood the  isothermal version of Eq.~(\ref{sum}) \citep[see Eq.~(15) of][]{makarieva20}.  \citet{er20} interpreted this relation as a contradiction in the reasoning of \citet{makarieva20}, since it can be concluded that with no latent heat input from the ocean there can be no storms, while dry hurricanes were shown to exist at least in numerical models \citep{mrowiec11,cronin19}. We note, however, that whatever follows from Eq.~(\ref{sum}), be that a contradiction or not, is an inherent feature of E-PI. All the equations that we have so far considered can be derived from E-PI's key equations, and vice versa, as we demonstrated in Section~\ref{frame}.

To elucidate the meaning of these relationships, let us consider yet another, equivalent, formulation
of Eq.~(\ref{sum}). The radial gradient of moist saturated entropy
can be also expressed in terms of $\pt T/\pt r$ and $\partial p/\partial r$ \citep[see][]{mn21}:
\begin{equation}
\label{dsT}
T \frac{\pt s^*}{\pt r} = -\alpha_d \frac{\pt p}{\pt r} \left( 1 + \frac{L \gamma_d^*}{RT} \right) \left(1 - \frac{1}{\Gamma}\frac{\pt T/\pt r}{\pt p/\pt r}\right),
\end{equation}
where $\Gamma$ is the moist adiabatic lapse rate (K~Pa$^{-1}$).
If, for simplicity, we neglect the water lifting by putting $K_1 = 0$, i.e., ignoring the term in square brackets in Eq.~(\ref{alg}),  we  notice, taking into account Eq.~(\ref{dsT}), that all the remaining terms in Eq.~(\ref{alg}) are proportional to the common factor $\alpha \pt p/\pt r$, which can be canceled. Again, putting for simplicity  $q_t= q^*  = 0$ and $\mathcal{B} = 1$, and using Eq.~(\ref{K2}), we obtain from Eq.~(\ref{alg}):
\beq\label{algfin}
1 = \varepsilon_C \left( 1 + \frac{L\gamma_d^*}{RT_{\rm b}} \right) \left(1 - \frac{1}{\Gamma}\frac{\pt T/\pt r}{\pt p/\pt r}\right) + \frac{v_{\rm c}}{r_{\rm c}}\frac{r_{\rm b}}{v_{\rm b}},
\eeq
which can be re-written in a form similar to Eq.~(\ref{E11}) using $\varepsilon_C = (T_{\rm b} - T_{\rm c})/T_{\rm b}$
and $\nu = v/r$:
\beq\label{A11}
\frac{\nu_{\rm b} - \nu_{\rm c}}{T_{\rm b} - T_{\rm c}} = \frac{\nu_{\rm b}}{T_{\rm b}} \left( 1 + \frac{L\gamma_d^*}{RT_{\rm b}} \right)  \left(1 - \frac{1}{\Gamma}\frac{\pt T/\pt r}{\pt p/\pt r}\right).
\eeq
In comparison to Eq.~(\ref{E11}), in this equation point $\rm b$ is not arbitrary but pertains to the point of maximum wind, while point $\rm c$ remains of an arbitrary choice.

Equations~(\ref{algfin}) and (\ref{A11}) show that when $\nu_{\rm c} = 0$ (no kinetic energy change in the outflow), and the air is horizontally isothermal, E-PI framework  presumes $\varepsilon_C (1 + L \gamma^*_d/RT_{\rm b}) = 1$ \citep{mn21}. This  relation between the water vapor mixing ratio and the outflow temperature has the following meaning.

As we already noted, in the E-PI framework, total work of the cycle is equal to the work on the path $\rm B'b$, where heat input occurs (see Eq.~(\ref{relm}) in Fig.~\ref{fig2}). This can only be the case when the adiabat $\rm bc$ is, at least somewhere, warmer than the adiabat $\rm B'C'$. Then the pressure deficit at $\rm b$ as compared to $\rm B'$ can be compensated by the pressure surplus in the free troposphere at $\rm bc$ as compared to $\rm B'C'$. Without this pressure surplus aloft, the work along $\rm bcC'B'$ will be negative rather than zero \citep[for a more detailed discussion, see][Fig.~1]{tellus17}.  When the air is horizontally isothermal, the required difference in the temperatures of the two adiabats can only be ensured by a higher water vapor mixing ratio $q^*$  and, accordingly, a higher partial pressure ratio $\gamma_d^*$, at $\rm bc$. How much of this water vapor condenses along $\rm bc$ producing the required temperature surplus, is dictated by the outflow temperature $T_{\rm c}$. The lower $T_{\rm c}$ (i.e., the higher $\varepsilon_C$), the lower $\gamma_d^*$ is required to ensure net zero work in the free troposphere.

When the cycle is dry, achieved by putting $L = 0$, then Eq.~(\ref{A11}), and the E-PI framework, lack a non-trivial solution for the isothermal case $\pt T/\pt r = 0$. 
Otherwise, total work of such a cycle would exceed that of a Carnot cycle, where total work is always lower than work on the warmer isotherm.  A dry Carnot cycle where total work is equal to the work at the warmer isotherm -- as it is in E-PI at the point of maximum wind -- is impossible.  A dry E-PI hurricane must have local air temperature increasing towards the center at the point of maximum wind.

\conclusions

\newdimen\bw
\bw=2.85cm
\newdimen\mh
\mh=1cm

\tikzstyle{blockr} = [rectangle, draw, fill=red!10, text width=\bw, text centered,minimum height=2.1cm]
\tikzstyle{lineA} = [->, >=open triangle 60,very thick,lightgray]
\tikzstyle{line} = [-, very thick,lightgray]
\tikzstyle{block1} = [rectangle, draw,text centered,text width=1\bw,minimum height=\mh]

\tikzstyle{blokeq} = [rectangle, draw=blue!10, very thick, text width=\bw, fill=blue!4, text centered, minimum height=1.5cm,execute at begin node=\setlength{\baselineskip}{12pt}]
\tikzstyle{blokeqs} = [rectangle, draw=none, text width=0.0\bw, text centered, minimum height=1.5cm,execute at begin node=\setlength{\baselineskip}{12pt}]
\tikzstyle{blokexpl} = [rectangle, very thick, draw = lightgray, fill=green!10, text centered, minimum height=1.5cm,execute at begin node=\setlength{\baselineskip}{12pt},rounded corners=8pt]

\tikzstyle{blokeq2} = [rectangle, draw=blue!10, very thick, text width=\bw, fill=blue!4, text centered, minimum height=1.7cm,execute at begin node=\setlength{\baselineskip}{12pt}]

\tikzstyle{comm} = [rectangle, draw,dotted,thick,fill=gray!2, text centered, minimum height=1.5cm,execute at begin node=\setlength{\baselineskip}{12pt}]

\newdimen\nodcv
\nodcv=0.5cm
\newdimen\nodch
\nodch=1cm

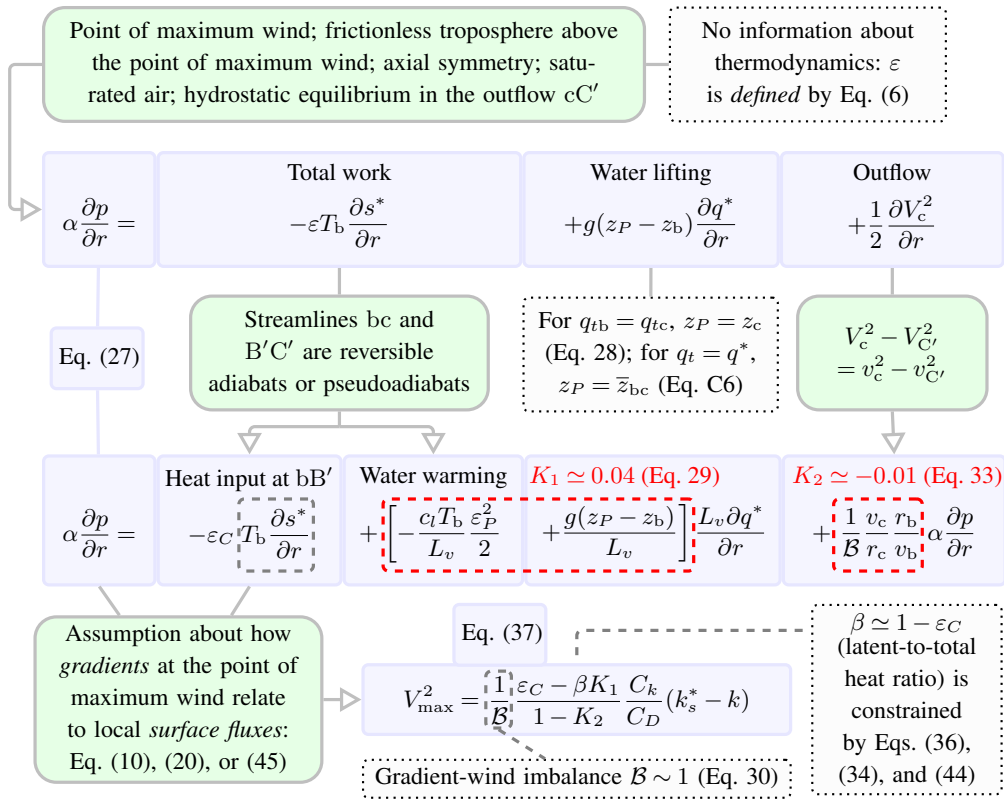
\begin{figure}[t]

\begin{tikzpicture}[node distance = \nodch and \nodcv, rounded corners=2pt]


\node[blokexpl, text width=4.18\bw,draw=none,fill=none](as1){};
\node[blokexpl, right=0cm of as1.west,node distance=0cm,text width=2.7\bw](as10){Point of maximum wind; frictionless troposphere above the point of maximum wind; axial symmetry; saturated air; hydrostatic equilibrium in the outflow $\rm cC'$};
\node[blokeqs,right=0.3cm of as10,draw,dotted,thick,fill=gray!2,text width = 1.2\bw](com1a){No information about thermodynamics: $\varepsilon$ is {\it defined}
by Eq.~(\ref{dQ})};

\node [below=1.8\nodch of as1.west, node distance=0cm](o){};
\node[blokeq, right=0cm of o.center,text width=0.44\bw, node distance=0cm](eq1) {$$\alpha \frac{\pt p}{\pt r} =$$};
\node[blokeq, right=0cm of eq1,text width=1.6\bw, node distance=0cm](eq1dQ) {$$-\varepsilon T_{\rm b}\frac{\pt s^*}{\pt r}$$};
\node[blokeq, right=0cm of eq1dQ, node distance=0cm,text width=1.1\bw](eq1wl) {$$+g(z_P - z_{\rm b})\frac{\pt q^*}{\pt r}$$};
\node[blokeq, right=0cm of eq1wl, node distance=0cm,text width=0.95\bw](eq1out) {$$+\frac{1}{2}\frac{\pt V_{\rm c}^2}{\pt r}$$};

\node [below=1.8\nodch of eq1wl.east, node distance=0cm](or){};
\node[blokeqs, left=-0.1cm of or.west,text width=1.1\bw,draw,dotted,thick,fill=gray!2](com1){For $q_{t \rm b} = q_{t \rm c}$, $z_P = z_{\rm c}$ (Eq.~\ref{def1});
for $q_t = q^*$, $z_P = \overline{z}_{\rm bc}$ (Eq.~\ref{Acalc})}; 
\draw[line](com1)--(eq1wl);
\draw[line](com1a)--(as10);

\node [below=2\nodch of eq1out.west](oo){};

\node[blokexpl, below=0.35\nodch of eq1out,text width=0.8\bw, node distance=0cm](as2){$V^2_{\rm c} - V^2_{\rm C'}$ $=v^2_{\rm c} - v^2_{\rm C'}$};

\node [below=2\nodch of eq1dQ.west](oo1){};

\node[blokexpl, below=0.35\nodch of eq1dQ,text width=1.3\bw](as3){Streamlines $\rm bc$ and $\rm B'C'$ are reversible adiabats or pseudoadiabats};

\node [below=5.9\nodch of as1.west, node distance=0cm](ob){};
\node[blokeq2, right=0cm of ob.center, node distance=0cm,text width=0.44\bw](eq2) {$$\alpha \frac{\pt p}{\pt r}=$$};
\node[blokeq2, right=0cm of eq2,text width=0.77\bw, node distance=0cm](eq2dQ) {$$- \varepsilon_C \,\,T_{\rm b}\frac{\pt s^*}{\pt r}$$};
\node[blokeq2, right=0cm of eq2dQ,text width=0.75\bw, node distance=0cm](eq2cl) {$$+\,\left[ -\frac{c_l T_{\rm b}}{L_v}\frac{\varepsilon^2_P}{2}\, \right.$$};
\node[blokeq2, right=0cm of eq2cl, node distance=0cm,text width=1.1\bw](eq2wl) {$$\left.+\frac{g(z_P - z_{\rm b})}{L_v} \right] \frac{L_v \pt q^*}{\pt r}$$};
\node[blokeq2, right=0cm of eq2wl, node distance=0cm,text width=0.95\bw](eq2out) {$$+\,\,\,\frac{1}{\mathcal{B}}\frac{v_{\rm c}}{r_{\rm c}}\frac{r_{\rm b}}{v_{\rm b}}\,\,\alpha \frac{\pt p}{\pt r}$$};

\node[below=0.08cm of eq2wl.east](eq2o){};
\node[blokeqs,right=0.55cm of eq2o.east,red,text width =0.335\bw,draw,dashed,very thick,minimum height=0.95cm](K2){};
\node[below=0.08cm of eq2cl.west](eq2clo){};
\node[blokeqs,right=0.4cm of eq2clo,red,text width =1.36\bw,draw,dashed,very thick,minimum height=0.95cm](K1){};
\node[above=0.65cm of K1.east](K1o){};\node[left=-0.55cm of K1o,red](K1n){$K_1 \simeq  0.04$~(Eq.~\ref{K1})};
\node[above=0.65cm of K2.west](K2o){};\node[right=-0.75cm of K2o,red](K2n){$K_2 \simeq -0.01$~(Eq.~\ref{K2})};

\node[below=0.08cm of eq2dQ.west](eq2dQo){};
\node[blokeqs,right=0.95cm of eq2dQo,text width =0.27\bw,draw,gray,dashed,very thick,minimum height=0.95cm](K1){};

\node[below=2.25\nodch of eq2.west](as4o){};
\node[blokexpl,right=0\bw of as4o.west,text width=1.25\bw](as4){Assumption about how {\it gradients} at the point of maximum wind relate to local {\it surface fluxes}: Eq.~(\ref{as1}),  (\ref{as2}), or (\ref{as3})};

\node[blokeq, right=0.5cm of as4, text width = 1.9\bw, minimum height = 0.9cm](res){\vspace{-0.3cm}
$$V_{\rm max}^2 =\, \frac{1}{\mathcal{B}}\,\frac{\varepsilon_C  - \beta K_1}{1 - K_2}\frac{C_k}{C_D}(k_s^* - k)$$}; 

\node[blokeqs,left=0.84cm of res.center,text width =0.044\bw,draw,dashed,gray,very thick,minimum height=0.75cm](Bcal){};
\node[comm,below=0.3cm of res,text width = 1.9\bw, minimum height=0.5cm](Bcal1){Gradient-wind imbalance $\mathcal{B}\sim 1$ (Eq.~\ref{Bsup})};
\draw[line,gray,very thick,dashed](Bcal.south)--(Bcal1);

\node[comm,right=0.25cm of res, minimum height=0.8cm,text width = 0.8\bw](beta){\vspace{-0.1cm}$\beta \simeq 1 -\varepsilon_C$ (latent-to-total heat ratio) is constrained by Eqs.~(\ref{algn}), (\ref{FV}), and (\ref{sum})};
\node[above=0.8cm of beta.west,node distance = 0cm](beta1){};
\draw[line,gray,very thick,dashed](beta1)-|(res.north);

\node[below=1.85cm of eq1.west](Eq25o){};
\node[blokeq, right=0.1cm of Eq25o.center,minimum height = 0.8cm,text width=0.35\bw](Eq25){Eq.~(\ref{alg})};
\draw[line,blue!10](Eq25) -- (eq2);
\draw[line,blue!10](Eq25) -- (eq1);

\node[left=0.9cm of res.north](Eq37o){};
\node[blokeq, above=0.065cm of Eq37o.south,minimum height = 0.8cm,text width=0.35\bw](Eq37){Eq.~(\ref{Vmaxn})};

\node[below=0.05cm of eq1out.north](out){Outflow};
\node[below=0.05cm of eq1dQ.north](dQn){Total work};
\node[below=0.05cm of eq1wl.north](wln){Water lifting};
\node[below=0.05cm of eq2dQ.north](dQ2){Heat input at $\rm bB'$};
\node[below=0.05cm of eq2cl.north](cl2){Water warming};

\node[below=0.2cm of as3.south, node distance=0cm](as3n) {};
\draw[line](as3.south) -- (as3n.north);
\draw[lineA](as3n.north) -| (eq2dQ.north);
\draw[lineA](as3n.north) -| (eq2cl.north);
\draw[line](as3.north) -- (eq1dQ.south);
\draw[line](eq1out.south) -- (as2.north);
\draw[lineA](as2.south) -> (eq2out.north);

\node[left=0.2cm of eq1.west, node distance=0cm](eq1w){};
\draw[lineA](as1.west)-|(eq1w.west) -> (eq1.west);
\draw[lineA](as4.east)->(res);

\draw[line](eq2.south) -- (as4);
\draw[line](eq2dQ.south) -- (as4);

\end{tikzpicture}
\caption{\label{fig3}
Main assumptions (green rounded boxes) and results (blue rectangular boxes) of this work with comments (dotted boxes).
}
\end{figure}

We considered an infinitely narrow steady-state thermodynamic cycle with the higher temperature corresponding
to the point of maximum wind, Figs.~\ref{fig1} and \ref{fig2}, and demonstrated its equivalence to the E-PI framework
as presented by \citet{em86} and \citet{emanuel11}. 
This revealed constraints not obvious in the original E-PI framework and clarified its physical meaning.  A summary of our results is given in Fig.~\ref{fig3}.
Since this analysis required many detailed derivations, we leave a comparably detailed discussion and development of the implications of 
these results to subsequent studies. Here we outline what we consider most essential. 

The water lifting term $g(z_P - z_{\rm b})$, with $z_P$ given by Eq.~(\ref{def1}), can be evaluated for any cycle with the known distribution 
of $q_t$ independent of the value of $\varepsilon$ (Fig.~\ref{fig3}). The E-PI thermodynamic cycle has a higher efficiency than the 
steady-state atmospheric circulation and tropical convection \citep[e.g.,][]{goody03},
which explains the less significant impact we estimated for E-PI storms (Table~\ref{EST}). 
Water lifting constitutes part of the total work of the cycle, which cannot be larger than 
$\varepsilon$ times the heat input. Since latent heat is a major part of heat input, its product with efficiency approximates total work. Both water lifting and latent heat input depend on the amount of evaporated water. Hence the ratio of the water lifting to total work is roughly equal to  the potential energy of precipitation $gz_P$ divided by the product of latent heat and efficiency, $K_1/\varepsilon \sim g z_P/(\varepsilon L_v)$.  With $K_1 \sim 10^{-2}$ and $\varepsilon \sim 10^{-2}$ the energy needed to lift water can exceed the 
total work of the cycle.

The infinitely narrow E-PI cycle is not a real steady-state cycle where evaporation equals precipitation, as the air does not descend along the adiabat $\rm C'B'$  (Fig.~\ref{fig1}). Here, instead, the water formally arises anew with its own non-zero gravitational energy. For this reason, the water lifting term in E-PI is proportional
to local evaporation ($\Delta q^*$) rather than precipitation ($q^*_{\rm b} - q^*_{\rm c}$). In real storms, precipitation $P$ at the point of maximum wind is significantly higher than
evaporation $E$. Replacement of $E$ with $P$ caused \citet{sabuwala15}'s overestimate of the water lifting term (Section~\ref{vmaxn}).

If the hurricane is composed of closed streamlines, each with a constant $q_t$ (the reversible case), the total work performed on lifting the water in such a storm is zero (Section~\ref{revh}).  However, since the E-PI cycle is not a cycle along which the air moves, the water lifting term here is higher in the reversible case than in the pseudoadiabatic case. 
This is due to the higher effective precipitation height in the former ($z_{\rm c} > a H_P$, Eq.~(\ref{zP})). Accounting for water warming,
which requires information about the cycle's thermodynamics (Fig.~\ref{fig3}), reduces this difference.  
For $T_{\rm b} \simeq 300$~K and the largest observed $\varepsilon_C \simeq 0.3$ \citep{demaria94}, the magnitudes of $K_1 \simeq 0.04$ (Eq.~\ref{K1}) 
and $\beta K_1/\varepsilon_C \simeq 0.1$ for reversible and pseudoadiabatic cases are similar. This corresponds to a $5\%$ reduction of $V_{\rm max}$ (Eq.~\ref{Vmaxn}). 
This reduction is larger for smaller $\varepsilon_C$ but smaller for lower $T_{\rm b}$ (Table~\ref{revnum}).
The developed analytical framework can be used to evaluate corresponding magnitudes for different scenarios in numerical models.

Our analysis clarifies that $\varepsilon$ in E-PI is not the actual efficiency of the cycle but the ratio of total work to heat input 
at $\rm B'b$ (Figs.~\ref{fig2} and \ref{fig3}). With additional heat inputs elsewhere in the cycle,
$\varepsilon$ in Eq.~(\ref{rel}) can be higher than Carnot efficiency (cf. Eq. \ref{dQd}).
This helps understand the phenomenon of "superintensity". When the adiabaticity is violated near the tropopause 
due to an extra heat input, E-PI's Eqs.~(\ref{alg}) and (\ref{Vmaxn}) can significantly underestimate $\alpha \pt p/\pt r$ and
the squared maximum velocity \citep[for details, see][]{mn21}. Since at the tropopause $q_t$ is approximately zero,
this extra heat input will not affect the $q_t$ distribution and, hence, will make the relative water lifting impact even smaller
(same absolute magnitude of $K_1$ related to greater total work).

Our derivations exposed the sensitivity of $V^2_{\rm max}$ to E-PI's key assumption about how the
gradients of respective variables at the point of maximum wind relate to their local surface fluxes.
If Eq.~(\ref{as2}) is applied, the extension of E-PI to unbalanced winds consists in dividing the conventional
E-PI squared velocity by factor $\mathcal{B}$ (\ref{Bsup}) that describes the deviation from gradient-wind balance \citep[for how 
this relates to the analysis of \citet{bryan09b}, see][]{mn21}. If, on the other
hand, Eq.~(\ref{as1}) is applied, the gradient-wind imbalance does not affect the E-PI formulation, see Eq.~(\ref{Vmaxn}).

The choice of the outflow point $\rm c$ as a point where $v_{\rm c} = 0$, i.e., putting $K_2 = 0$, is equivalent to postulating  that, for a cycle including the point of maximum wind, net work in the free troposphere is zero (see Eq.~(\ref{relm}) in Fig.~\ref{fig2}).  Since generally in the free troposphere the outflowing air has to move against the inward-pulling horizontal pressure gradient, compensating for this negative work requires extra warming at $\rm bc$ compared to $\rm C'B'$. This extra warming is provided either by a higher mixing ratio at $\rm b$ compared to $\rm B'$, or by a higher temperature, or by both.  This constraint takes the form of the dependence between the outflow temperature, the mixing ratio and 
the ratio of the horizontal gradients of temperature and pressure, see Eqs.~(\ref{algfin}) and (\ref{A11}).
It follows that attempts to retrieve total pressure fall from E-PI by assuming $\pt T/\pt r = 0$ cannot yield correct results under the conventional assumption of $K_2 =0$ \citep[cf.][p.~588]{em86}.

When the horizontal gradient of air temperature is moist adiabatic (which corresponds to zero heat input),
Eq.~(\ref{algfin}) reduces to $v_{\rm b}/r_{\rm b} = v_{\rm c}/r_{\rm c}$. This constancy of angular velocity
combined with angular momentum conservation along $\rm bc$ gives two solutions (see Eq.~\ref{K2}). One is trivial, $r_{\rm b} = r_{\rm c}$. Another one is $\nu_{\rm b} = \nu_{\rm c} =-f/2$; it describes an atmosphere at rest in the inertial frame of reference. In either case, there is no storm. However, we now know that, at least in models, it is possible to have a tropical cyclone with zero heat input from the ocean \citep{kieu20}, although further tapering with the conventional model parameters might be required to make such a cyclone more stable.
This prompts reconsidering the relevance of the local approach for the determination of maximum potential intensity.

A non-local constraint on the work in the free troposphere resulting from E-PI can be applied to the integral cycle $\rm ABCDA$. In this case, we cannot put $\pt V^2/\pt r = 0$ in Eq.~(\ref{relm}). The work in the free troposphere is equal to the non-zero increment of the kinetic energy in the boundary layer. 
Having reached the eyewall, the air must then have sufficient energy to flow away from the hurricane. If not generated in the boundary layer, this energy could derive from a pressure gradient in the upper atmosphere:  if at the expense of the hurricane's extra warmth  the air pressure in the column above the area of maximum wind is higher than in the ambient environment, this pressure gradient will accelerate the air outward. 
However, a significant pressure deficit at the surface precludes the formation of a significant pressure surplus aloft \citep[e.g.,][Fig.~1d]{tellus17}.

Moreover, this pressure deficit is what accelerates air in the boundary layer. If the pressure gradient is sufficiently steep and the radial motion sufficiently rapid, air expansion will be accompanied by a drop of temperature. The process is closer to an adiabat than to an isotherm, as it was, for example, in Hurricane Isabel 2003 \citep{montgomery06,aberson06,mn21}.  As the warm air creates a pressure surplus aloft facilitating the outflow, cold air creates a pressure deficit. This enhances the pressure gradient in the upper atmosphere against which the air must work to leave the hurricane. Consequently, the storm cannot deepen indefinitely. Eventually, {\it the kinetic energy acquired in the boundary layer} becomes insufficient  for the rising and adiabatically cooling air to overcome the pressure gradient in the upper atmosphere, and the outflow must weaken. This condition could provide distinct constraints on storm intensity. Further research is needed to see whether such processes are relevant in real storms.

\section*{Acknowledgments}
We are grateful to three anonymous referees for their useful comments. Our response to them can be found in the Appendix. Work of A.M. Makarieva is partially funded by the Federal Ministry of Education and Research (BMBF) and the Free State of Bavaria under the Excellence Strategy of the Federal Government and the L\"ander, as well as by the Technical University of Munich -- Institute for Advanced Study.


\section{Appendices}
\renewcommand{\thesubsection}{Appendix~\Alph{subsection}}%

\setcounter{equation}{0}%
\setcounter{table}{0}%
\renewcommand{\theequation}{A\arabic{equation}}%
\renewcommand{\thetable}{A\arabic{table}}%

\subsection{Deriving Eq.~(\ref{Fdl}) from the equations of motion}
\label{A0}

One of our reviewers stated that Eq.~\eqref{Fdl}, stemming from the Bernoulli equation, could not be obtained assuming hydrostatic equilibrium \eqref{he} alone, but additionally requires the gradient-wind balance. To show that it is not the case and to facilitate a comparison of our approach with the available studies \citep[e.g.,][]{rotunno12}, we follow the reviewer's suggestion  to derive the Bernoulli equation from  the equations of motion. These equations in a reference frame rotating with angular velocity $\bm{\Omega}$ can be cast into the vector form as follows  \citep[see, e.g.,][]{lorenz67,vallis06}
\beq\label{AA1}
 \frac{d  \mathbf{V}}{dt} =  - \alpha \nabla p - 2 [\bm{\Omega}\times \mathbf{V}] - \nabla \Phi +  \mathbf{F}.
\eeq
Here, $\mathbf{V}$ is the total velocity of air motion, relative to rotating Earth; $\alpha \equiv 1/\rho$ is the specific volume; 
$p$ is air pressure.  On the right-hand side of Eq.~\eqref{AA1}, the first term is the  pressure-gradient force acting on an air parcel of unit mass from the side of its surrounding air; the second term describes the Coriolis acceleration; $\mathbf{F}$ is frictional force per unit mass. The geopotential $\Phi$ is defined such that $\mathbf{g}= - \nabla \Phi$, where $\mathbf{g}$ is the  effective gravity, which in addition to the acceleration due to gravity also takes into account the centrifugal acceleration.

The material derivative is given by
\beq
\frac{d}{dt} = \frac{\partial }{\partial t} + (\mathbf{V} \cdot \nabla ) .
\eeq
Using the relation $ (\mathbf{V} \cdot \nabla ) \mathbf{V}= \nabla V^2/2 - \mathbf{V} \times [\nabla  \times \mathbf{V}]$, one can write  Eq.~\eqref{AA1} for the case of a steady state ($ \partial \mathbf{V}/\partial t =0$) as
\beq\label{AA2}
 - \alpha \nabla p = \frac12 \nabla V^2 + [\bm{\omega} \times  \mathbf{V}] - \mathbf{g}- \mathbf{F}, 
\eeq
where $\bm{\omega} \equiv  [\nabla  \times \mathbf{V}] + 2 \bm{\Omega}$ is the absolute vorticity. 

Let us find the inner product of Eq.~\eqref{AA2} and vector $d\mathbf{l}$. The equality
\beq\label{AA3}
[\bm{\omega}  \times \mathbf{V}] \cdot d\mathbf{l} = 0
\eeq
holds if $d\mathbf{l}$ is the length element along a streamline, i.e., by definition is tangent to the local velocity $\mathbf{V}= d\mathbf{l}/dt$. Indeed, Eq.~\eqref{AA3} follows from the relations
\beq
[\bm{\omega}  \times \mathbf{V}] \cdot \mathbf{V} = \bm{\omega}  \cdot  [\mathbf{V} \times \mathbf{V}] = 0 .
\eeq
As a result, one obtains the Bernoulli equation
\beq\label{AA2b}
 - \alpha \nabla p \cdot d\mathbf{l} = \frac12 \nabla V^2 \cdot d\mathbf{l} - \mathbf{g} \cdot d\mathbf{l} - \mathbf{F} \cdot d\mathbf{l} ,
\eeq
which is valid along a streamline.  The meaning of Eq.~\eqref{AA3} is that the force  $[\bm{\omega}  \times \mathbf{V}]$ (per unit mass) caused by the vorticity (including the Coriolis force) does not perform work and, accordingly, it does not change the energy of moving air parcel. 

If the air motion possesses axial symmetry, it is convenient to use the cylindrical coordinate system with the basis vectors $\mathbf{e}_r$,  $\mathbf{e}_\theta$, and $\mathbf{e}_z$. The position vector of a point $\mathbf{r}$ is now characterized by three components, namely, by  radial distance $r$, azimuth $\theta$, and  height $z$. The length element  of  a streamline and the gradient operator read
\begin{align}
d\mathbf{l} & = dr \mathbf{e}_r + r d\theta \mathbf{e}_\theta + dz \mathbf{e}_z  = (u \mathbf{e}_r + v \mathbf{e}_\theta + w \mathbf{e}_z) dt= \mathbf{V} dt , \label{AA4}\\
\nabla &\equiv \frac{\partial}{\partial \mathbf{l}} = \mathbf{e}_r \frac{\partial }{\partial r}+\mathbf{e}_\theta \frac{1}{r}\frac{\partial }{\partial \theta}+ \mathbf{e}_z \frac{\partial }{\partial z} .
\end{align}
The curl of velocity (relative vorticity) may be written in the form of a determinant
\begin{equation}
\mathrm{curl}~\mathbf{V}  \equiv [\nabla \times \mathbf{V}] = \frac{1}{r}
\begin{vmatrix} \mathbf{e}_r & r \mathbf{e}_\theta & \mathbf{e}_z   \\ 
\partial/\partial r & \partial/\partial \theta & \partial/\partial z \\
u & rv & w \end{vmatrix} = \left( \frac{1}{r} \frac{\partial w}{\partial \theta} - \frac{\partial v}{\partial z}  \right)  \mathbf{e}_r 
+  \left( \frac{\partial u}{\partial z} - \frac{\partial w}{\partial r}  \right)  \mathbf{e}_\theta +  \frac{1}{r} \left(  \frac{\partial (rv)}{\partial r} - \frac{\partial u}{\partial \theta}  \right)  \mathbf{e}_z.  
\end{equation}

The geopotential is given by  $\Phi = g z $, so that $\mathbf{g}= - \nabla \Phi = - g  \mathbf{e}_z $. It is usual to assume that the contribution of the Coriolis force to the vertical ($z$) component of the equations of motion is small with respect to the contribution of the centrifugal force and can be neglected. Then  $\bm{\omega} \simeq [\nabla  \times \mathbf{V}] + f \mathbf{e}_z$, where $f= 2\Omega \sin \varphi$ is the Coriolis parameter ($\Omega = 2\pi/T$ is the rotation rate of the Earth, $T=24$~h  is  the rotation period of the Earth  and $\varphi$ is latitude).

For  axisymmetric motion, $p$ and $\mathbf{V}$ are independent of  the angle $\theta$. In this case, the cylindrical components of  Eq.~\eqref{AA2} have the form
\begin{subequations}\label{A10}
\begin{align}\label{m1}
- \alpha \frac{\partial p}{\partial r} &= \frac{1}{2} \frac{\partial V^2}{\partial r} + w \omega_\theta - v \omega_z - F_r , \\ \label{m2}
0 &= \hphantom{\frac{1}{2} \frac{\partial V^2}{\partial r}} + u \omega_z - w  \omega_r - F_\theta , \\ \label{m3}
- \alpha \frac{\partial p}{\partial z} &= \frac{1}{2} \frac{\partial V^2}{\partial z} -u \omega_\theta + v \omega_r +g - F_z , 
\end{align}
\end{subequations}
where $\omega_r = - \partial v/\partial z$, $ \omega_\theta = \partial u/\partial z - \partial w/\partial r$ and $ \omega_z= f + r^{-1}\partial (r v)/\partial r$.  To derive the Bernoulli equation, we consider the streamline defined by $d\mathbf{l}$, Eq.~\eqref{AA4}. Multiplying the above three equations by $dr$, $rd\theta$ and $dz$, respectively,  and summing them  one obtains 
\begin{equation}
\label{B1}
 - \alpha \frac{\partial p}{\partial \mathbf{l}}\cdot d\mathbf{l} =\frac{1}{2} \frac{\partial V^2}{\partial \mathbf{l}}\cdot d\mathbf{l} + g dz - \mathbf{F}\cdot  d\mathbf{l} ,
\end{equation}
where $\partial p/\partial \theta =0$ and $\partial \mathbf{V}/\partial \theta =0$. The cancellation of all terms involving the absolute vorticity components is a consequence of the more general formula~\eqref{AA3}.

Applying hydrostatic equilibrium (\ref{he}) and the condition $\pt V/\pt z = 0$ turns Eq.~(\ref{m3}) into $-u \omega_\theta + v \omega_r  - F_z=0$ and the Bernoulli equation (\ref{B1}) into
\beq\label{B2}
 - \alpha \frac{\partial p}{\partial r}dr =\frac{1}{2} \frac{\partial V^2}{\partial r}dr - \mathbf{F}\cdot  d\mathbf{l},
\eeq
which yields our Eq.~(\ref{Fdl}).

\setcounter{equation}{0}%
\setcounter{table}{0}%
\renewcommand{\theequation}{B\arabic{equation}}%
\renewcommand{\thetable}{B\arabic{table}}%

\subsection{Extra work due to warming precipitating water: Deriving Eq.~(\ref{stc})}
\label{eh}

Saturated moist entropy per unit dry air mass is defined as \citep[see][Eq.~(A4)]{pa11}
\beq\label{s}
s^* = (c_{pd} + q_t c_l) \ln \frac{T}{T_0} - \frac{R}{M_d} \ln \frac{p_d}{p_0} + q^* \frac{L_v}{T} .
\eeq
Here,  $L_v = L_{v0} + (c_{pv} - c_l)(T-T_0)$ is the latent heat of vaporization (J~kg$^{-1}$); $q^* = \rho^*_v/\rho_d$,  $q_l = \rho_l/\rho_d$, and  $q_t = q^* + q_l$ are the mixing ratio for saturated water vapor, liquid water, and total water, respectively; $\rho_d$, $\rho^*_v$, and $\rho_l$ are the density of dry air, saturated water vapor and liquid water, respectively;  $c_{pd}$ and $c_{pv}$ are the specific heat capacities of dry air and water vapor at constant pressure; $c_l$ is the specific heat capacity of liquid water; $R = 8.3$~J~mol$^{-1}$~K$^{-1}$ is the universal gas constant; $M_d$ is the molar mass of dry air;  $p_d$ is the partial pressure of dry air; $T$ is the temperature; $p_0$ and $T_0$ are reference air pressure and temperature.

The ideal gas law for the partial pressure $p_v = p - p_d$ of  water vapor is
\beq\label{igv}
p_v = N_v RT,  \quad N_v = \frac{\rho_v}{M_v} , 
\eeq
where  $M_v$ and $\rho_v$ are the molar mass and density of water vapor, respectively. Using Eq.~(\ref{igv}) with $p_v = p_v^*$ in the definition of $q^*$ 
\beq\label{q}
q^* \equiv \frac{\rho^*_v}{\rho_d} = \frac{M_v}{M_d} \frac{p_v^*}{p_d} \equiv \frac{M_v}{M_d} \gamma_d^*,\quad 
\gamma_d^* \equiv\frac{p_v^*}{p_d},
\eeq
and  applying the Clausius-Clapeyron law 
\beq\label{CC}
\frac{dp_v^*}{p_v^*} = \xi \frac{dT}{T}, \quad L \equiv L_v M_v , \quad \xi \equiv \frac{L}{RT} ,
\eeq
we obtain from Eq.~(\ref{s})
\begin{subequations}\label{Tds}
\begin{align}
T ds^* & = d(L_v q^*) - \alpha_d dp +  (c_{pd} + c_l q_t) dT + c_l T \ln \frac{T}{T_0} dq_t  \label{Tds:a}\\
& = L_v dq^* - \alpha_d dp +  c_{p} dT +  c_l T \ln \frac{T}{T_0} dq_t ,  \label{Tds:b}
\end{align}
\end{subequations}
where $c_{p} \equiv c_{pd} + q^*c_{pv} + q_{l} c_{l }$ and $\alpha_d \equiv  1/\rho_d$.
Integrating Eq.~(\ref{Tds:a}) with $T_0 = T_{\rm b}$ yields
\begin{equation}\label{oints}
-\oint \alpha_d dp = \oint T ds^* - c_l \oint q_t dT -c_l \oint T \ln \frac{T}{T_{\rm b}} dq_t  .
\end{equation}
Along the isotherms $dT = 0$, i.e., the paths $\rm B'b$ and $\rm cC'$ are characterized by temperatures 
$T_{\rm b}$ and  $T_{\rm c}$, respectively, while $ds^* = 0$ along the adiabats $\rm bc$ and $\rm C'B'$, i.e.,  $s^*_{\rm b} = s^*_{\rm c}$, $s^*_{\rm B'} =  s^*_{\rm C'}$
both for reversible ($dq_t = 0$) and pseudoadiabatic ($q^* = q_t$) cases.
The first term on the right-hand side of Eq.~\eqref{oints} can be written as follows
\begin{equation}\label{term-1}
\oint T ds^* = \int \limits^{\rm b}_{\rm B'} T ds^* + \int \limits^{\rm C'}_{\rm c} T ds^* =
\left( 1 - \frac{T_{\rm c}}{T_{\rm b}}\right) \int_{\rm B'}^{\rm b} T ds^*.
\end{equation}
Here we took into account the fact that  integral over the colder isotherm can be expressed in terms of  integral over the warmer isotherm \citep{pa11}:
\begin{equation}
\label{dq}
\int \limits_{\rm c}^{\rm C'} T ds^* = T_{\rm c}\int \limits_{\rm c}^{\rm C'} ds^* = T_{\rm c}s^* \bigg|_{\rm c}^{\rm C'} = T_{\rm c}(s^*_{\rm B'} -  s^*_{\rm b}) = - \frac{ T_{\rm c}}{ T_{\rm b}} \int \limits^{\rm b}_{\rm B'} T ds^* .
\end{equation}

In the reversible case (i.e., $q_{t\rm b} = q_{t\rm c}$ and $q_{t\rm B'}   = q_{t\rm C'}$), the second and third circulation integrals on the right-hand side of Eq.~\eqref{oints} are evaluated as follows
\begin{gather}
\oint q_t dT = -\oint T  dq_t =\int \limits^{\rm c}_{\rm C'} T dq_t -\int \limits^{\rm b}_{\rm B'} T dq_t  =  \Delta  q_{t} (T_{\rm c} - T_{\rm b}) , \label{term-2}\\
\oint T \ln \frac{T}{T_{\rm b}} dq_t = T_{\rm c}\ln\frac{T_{\rm c}}{T_{\rm b}} \int  \limits_{\rm c}^{\rm C'}dq_t = - \Delta  q_{t}T_{\rm c} \ln\frac{T_{\rm c}}{T_{\rm b}}  , \label{term-3}
\end{gather}
where $\Delta  q_{t} \equiv q_{t\rm b} - q_{t\rm B'} = q_{t\rm c} - q_{t\rm C'}$. 

Using Eqs.~\eqref{term-1}, \eqref{term-2} and \eqref{term-3}, integral~\eqref{oints} can be cast into the following form
\begin{align}\label{oints1}
-\oint \alpha_d dp & = \left( 1 - \frac{T_{\rm c}}{T_{\rm b}}\right)\int_{\rm B'}^{\rm b} Tds^* + c_l \Delta q_t \left[T_{\rm b} - T_{\rm c} +  T_{\rm c} \ln \frac{T_{\rm c}}{T_{\rm b}}\right]  \nonumber\\
&=\varepsilon_C \int_{\rm B'}^{\rm b} Tds^* + c_l T_{\rm b} \Delta q_t \left[\varepsilon_C + (1 - \varepsilon_C) \ln (1 - \varepsilon_C) \right]. 
\end{align}
Taking into account that for $\varepsilon_C =  1-T_{\rm c}/T_{\rm b}\lesssim 1$
\beq\label{varep}
\varepsilon_C + (1 - \varepsilon_C) \ln (1 - \varepsilon_C) \simeq \frac{\varepsilon_C^2}{2},
\eeq
we obtain Eq.~(\ref{stc}) from Eq.~(\ref{oints1}). For the largest observed $\varepsilon_C = 0.35$, approximation (\ref{varep}) is a $13\%$ underestimate.

For an infinitely narrow cycle, with $\rm B' \to  b$, Eqs.~(\ref{term-1}) and  (\ref{term-2})--(\ref{oints1}) are valid even if $\rm B'b$ and $\rm cC'$ are {\it not} isotherms, due to the smallness of temperature change on these paths as compared to the finite difference $T_{\rm b} - T_{\rm c}$.

In the pseudoadiabatic case (i.e., $q_t = q^*$), the second and third circulation integrals on the right-hand side of Eq.~\eqref{oints} are evaluated along the isothermal interval ${\rm B'b}$ and unclosed contour ${\rm bcC'B'}$
(denoted by symbol $\curvearrowright$):
\begin{gather}
-\oint q_t dT= \oint T dq^*=  \int \limits^{\rm b}_{\rm B'} T dq^* +  \int\limits_{\curvearrowright}  Tdq^*= \Delta q^* (T_{\rm b}- T_{P}), \label{PAEH-1}\\
- \oint T\ln\frac{T}{T_{\rm b}} dq_t= - \int\limits_{\curvearrowright}  T\ln\frac{T}{T_{\rm b}} dq^* \approx 
-  \ln\frac{T_{P}}{T_{\rm b}} \int\limits_{\curvearrowright} T dq^* =  \Delta q^* T_{P}\ln\frac{T_{P}}{T_{\rm b}} . \label{PAEH-2}
\end{gather}
Here $\Delta  q^* \equiv q^*_{\rm b} - q^*_{\rm B'}$, $T_P$ defined in Eq.~(\ref{def1}) designates the mean temperature at which water vapor  condenses and precipitates. In Eq.~\eqref{PAEH-2} the first equality implies that the integral over the warmer isotherm gives zero contribution. Then we estimate the integral by assuming that $\ln T/T_{\rm b}$ is a slowly varying function with respect to $T$, so that it can be taken out of the integral at the point $T= T_P$. 
With the use of Eqs.~\eqref{PAEH-1} and \eqref{PAEH-2} the water warming term in the pseudoadiabatic case is	
\begin{equation}
\label{clpseudo}
c_l \Delta q^* \left(T_{\rm b} - T_P + T_P \ln \frac{T_P}{T_{\rm b} } \right) = c_l T_{\rm b}\Delta q^*
\left[ \varepsilon_P + (1 -\varepsilon_P) \ln (1- \varepsilon_P) \right] \simeq c_l T_{\rm b}\Delta q^*
 \frac{\varepsilon_P^2}{2},
\end{equation}
where $\varepsilon_P \equiv 1 - T_P/T_{\rm b}$. We discuss the accuracy of this expression in the end of the next section.

To obtain Eq.~(\ref{dsq}), we use Eqs.~(\ref{q}) and (\ref{CC}) to exclude the temperature differential from Eq.~(\ref{Tds}):
\begin{align}
\label{ddq}
\frac{dq^*}{q^*} & = \frac{d\gamma^*_d}{\gamma^*_d} = (1 + \gamma^*_d) \left(\frac{dp_v^*}{p_v^*} - \frac{dp}{p} \right) = (1 + \gamma^*_d) \left(\xi \frac{dT}{T} - \frac{dp}{p} \right),\\
\label{dT}
dT & = \left(\frac{dq^*}{q^*} + \frac{dp}{p_d}\right) \frac{T}{\xi (1 + \gamma^*_d)} = \left(\frac{dq^*}{q^*} +  \alpha_d  \frac{M_d}{RT} dp \right) \frac{T}{\xi (1 + \gamma^*_d)}.
\end{align}
Putting Eq.~(\ref{dT}) into Eq.~(\ref{Tds:b}) with $T_0 = T$ and grouping the $dq^*$ and $dp$ terms yields Eq.~(\ref{dsq}).
Combining Eqs.~(\ref{algn}), (\ref{FV}) and (\ref{sum}) we obtain for $\beta$ in Eq.~(\ref{algn})
\begin{equation}\label{beta}
\beta =\frac{1 -\varepsilon_C(1-\varkappa_2) - K_2}{(1-K_2)
(1+\varkappa_1) - K_1(1-\varkappa_2)}.
\end{equation}

\vspace{0.2cm}

\setcounter{equation}{0}%
\setcounter{table}{0}%
\renewcommand{\theequation}{C\arabic{equation}}%
\renewcommand{\thetable}{C\arabic{table}}%

\subsection{Water lifting: Deriving Eq.~(\ref{alg})}
\label{wl}

In Eq.~(\ref{est0}), we represent the last but one term on the right-hand side as
\begin{subequations} \label{est1a}
\begin{align}
\oint \alpha q_t dp &=  \int_{\rm B'}^{\rm b} \alpha q^* \frac{\partial p}{\partial r}dr - \int_{\rm b}^{\rm c} q_t d\left(\frac{V^2}{2} + gz\right) - \int_{\rm c}^{\rm C'} q_t g dz - \int_{\rm C'}^{\rm B'} q_t d\left(\frac{V^2}{2} + gz\right)  \label{est1a:a}\\ 
&= \int_{\rm B'}^{\rm b} \alpha q^* \frac{\partial p}{\partial r}dr - q_t\frac{V^2}{2}\bigg|_{\rm b}^{\rm c} - q_t\frac{V^2}{2}\bigg|_{\rm C'}^{\rm B'} +\int_{\rm b}^{\rm c} \frac{V^2}{2}dq_t  +\int_{\rm C'}^{\rm B'} \frac{V^2}{2}dq_t  -\int_{\curvearrowright} gq_tdz \\
&= \int_{\rm B'}^{\rm b} \alpha q^* \frac{\partial p}{\partial r}dr + q^*\frac{V^2}{2}\bigg|_{\rm B'}^{\rm b} + q_t\frac{V^2}{2}\bigg|_{\rm c}^{\rm C'} +\int_{\rm b}^{\rm c} \frac{V^2}{2}dq_t  +\int_{\rm C'}^{\rm B'} \frac{V^2}{2}dq_t -  \int_{\curvearrowright} gq_tdz \\
&= \int_{\rm B'}^{\rm b} q^* \left( \alpha \frac{\partial p}{\partial r} + \frac{1}{2}\frac{\partial V^2}{\partial r} \right) dr +\int_{\rm B'}^{\rm b} \frac{V^2}{2}\frac{\partial q^*}{\partial r} dr  + q_t\frac{V^2}{2}\bigg|_{\rm c}^{\rm C'} +\int_{\rm b}^{\rm c} \frac{V^2}{2}dq_t  +\int_{\rm C'}^{\rm B'} \frac{V^2}{2}dq_t -  \int_{\curvearrowright} gq_tdz \\
&= \int_{\rm B'}^{\rm b} q^* \left( \alpha \frac{\partial p}{\partial r} + \frac{1}{2}\frac{\partial V^2}{\partial r} \right) dr +\int_{\rm B'}^{\rm b} \frac{V^2}{2}\frac{\partial q^*}{\partial r} dr + \int_{\rm c}^{\rm C'}\frac{q_t }{2}\frac{\partial V^2}{\partial z} dz  +\int_{\curvearrowright} \frac{V^2}{2}dq_t - \int_{\curvearrowright} gq_tdz.   \label{est1a:e}
\end{align}
\end{subequations}
Here it is assumed that $q_l =0$ at the isotherm ${\rm B'b}$. In Eq.~\eqref{est1a:a}, we have used the Bernoulli equation for the two streamlines, $\rm bc$ and $\rm B'C'$, and the hydrostatic equilibrium equation (\ref{he}) for the vertical path $\rm cC'$. The unclosed contour ${\rm bcC'B'}$ is denoted by symbol  $\curvearrowright$. Note that $z_{\rm B'} = z_{\rm b}$ and $r_{\rm c} = r_{\rm C'}$.

Writing the last integral in  Eq.~\eqref{est1a:e} as
\begin{equation}\label{last}
\int_{\curvearrowright} q_t dz = q^*z\bigg|_{\rm b}^{\rm B'} - \int_{\curvearrowright}  z dq_t \equiv  (z_P - z_{\rm b})\Delta q^*,
\end{equation}
we obtain 
\begin{equation}\label{qdp}
\oint \alpha q_t dp = -\int_{\rm b}^{\rm B'} q^* \left( \alpha \frac{\partial p}{\partial r} + \frac{1}{2}\frac{\partial V^2}{\partial r} \right) dr -\int_{\rm b}^{\rm B'} \frac{V^2}{2}\frac{\partial q^*}{\partial r} dr  -  \left[g(z_P - z_{\rm b}) + \frac{V^2_P}{2}\right]\Delta q^* + \int_{\rm c}^{\rm C'} \frac{q_t}{2}\frac{\partial V^2}{\partial z} dz.
\end{equation}
Here $\Delta q^* \equiv q^*_{\rm b} - q^*_{\rm B'}$ and $z_P$ and $V_P^2$ are defined in Eq.~(\ref{def1}).

Putting Eq.~(\ref{qdp}) into Eq.~(\ref{est0}) yields
\begin{multline}\label{prefinal}
\int_{\rm b}^{\rm B'} (1 + q^*) \left( \alpha \frac{\partial p}{\partial r} + \frac{1}{2}\frac{\partial V^2}{\partial r} \right)dr  = -\varepsilon_C \int_{\rm b}^{\rm B'} T \frac{\partial s^*}{\partial r} dr - c_l T_{\rm b}\frac{\varepsilon^2_P}{2}\int_{\rm b}^{\rm B'} \frac{\partial q^*}{\partial r} dr\\
-\int_{\rm b}^{\rm B'} \frac{V^2}{2}\frac{\partial q^*}{\partial r} dr  -\left[g (z_P - z_{\rm b}) + \frac{V^2_P}{2}\right]\Delta q^*  +\int^{\rm C'}_{\rm c} (1 + q_t) \frac{1}{2}\frac{\partial V^2}{\partial z}dz.
\end{multline}
In the limit $r_{\rm B'} \to r_{\rm b}$ and $z_{\rm C'} \to z_{\rm c}$ we can lift the integral signs in Eq.~(\ref{prefinal}) and divide both parts of the equation by $dr$ to obtain Eq.~(\ref{alg}). Note that by chain rule $\left(\partial V_{\rm c}/\partial z\right) dz|_{r=r_{\rm c}}= \left(\partial V_{\rm c}/\partial r\right) dr|_{z=z_{\rm b}}$.

To evaluate $V_P^2$, $T_P$ and $z_P$ for the case $q_t = q^*$, we can express the corresponding definitions (\ref{def1}) as follows:
\begin{subequations} \label{A}
\begin{align}
X_P \Delta q^* & \equiv  -\int_{\curvearrowright} X dq^* = \overline{X}_{\rm bc}(q^*_{\rm b} - q^*_{\rm c}) + 
\overline{X}_{\rm cC'} (q^*_{\rm c} - q^*_{\rm C'}) + \overline{X}_{\rm C'B'}(q^*_{\rm C'} - q^*_{\rm B'}), \label{Adef:a}\\ 
\overline{X}_{\rm kj} &\equiv -\frac{1}{q^*_{\rm k} - q^*_{ \rm j}}\int_{\rm k}^{\rm j} X dq^*, \label{Adef}
\end{align}
\end{subequations}
where $\Delta q^* \equiv q^*_{\rm b} - q^*_{\rm B'}$, $X = \{V^2, T, z \}$ and $\overline{X}_{\rm kj}$ is the average
value of $X$ over path $\rm kj$.

In the limit $r_{\rm B'} \to r_{\rm b}$ and $z_{\rm C'} \to z_{\rm c}$, when the two streamlines coincide, we have from Eq.~(\ref{A})
\beq\label{Acalc}
X_P =\overline{X}_{\rm bc} + \frac{\pt \overline{X}_{\rm bc}}{\pt q^*_{\rm
b}} (q^*_{\rm b} - q^*_{\rm c}) + (X_{\rm c} - \overline{X}_{\rm bc}) \frac{q^*_{\rm c} - q^*_{\rm C'}}{q^*_{\rm b} - q^*_{\rm B'}} ,
\eeq
where $X_{\rm c} = \overline{X}_{\rm cC'}$.  When $(q^*_{\rm c} - q^*_{\rm C'})/(q^*_{\rm b} - q^*_{\rm B'}) \sim q^*_{\rm c}/q^*_{\rm b} \ll 1$ (which is always the case when $z_{\rm c}$ is sufficiently large), the last term on the right-hand part of Eq.~(\ref{Acalc}) can be neglected.

Due to the Clausius-Clapeyron law (\ref{CC}), the relative change of temperature is about $\xi^{-1} \sim 0.05$ of the relative change of
$p_v^*$ and, hence, of $q^*$: $q^*_{\rm b} \pt \overline{T}_{\rm bc}/\pt q^*_{\rm b} \sim  \overline{T}_{\rm  bc}/\xi$. Therefore, for $X=T$ the second term on the right-hand side of  Eq.~(\ref{Acalc}) can be neglected, and $T_P =  \overline{T}_{\rm
bc}$, which is the mean temperature of precipitation along $\rm bc$.

With the mean temperature of precipitation approximately independent of $q^*_{\rm b}$, the mean precipitation height $H_P \equiv
\overline{z}_{\rm bc}$ scales as $H_P \sim T_P/\Gamma \sim q^*_{\rm b}$, because for sufficiently large $q^*$ the moist adiabatic lapse rate $\Gamma$ is approximately inversely proportional to $q^*$ \citep[e.g.,][Eq.~(A10)]{mn21}. Therefore, for $X = z$ the second term in Eq.~(\ref{Acalc}) is approximately equal to the first one, such that
\beq\label{zP}
z_P = a H_P
\eeq
with $a \sim 2$.  For more accurate estimates, we evaluated hydrostatic saturated reversible and pseudoadiabatic profiles ($ds^* = 0$)
with surface pressure $p_s = 950$~hPa, $z_{\rm b} = 1$~km and variable $T_{\rm b}$ and $z_{\rm c}$ (Table~\ref{revnum}).
For the same $z_{\rm c}$, the values of $T_{\rm c}$ are different, since the reversible moist adiabatic lapse rate is smaller than the pseudoadiabatic one \citep[e.g.,][Fig.~1a]{mapes01}. The dependence of $H_P$ on $q^*$ in Eq.~(\ref{Acalc}) was evaluated by varying $p_s$ at constant surface temperature $T_s$. For $T_{\rm b} = 300$~K and $z_{\rm c} = 18$~km we have $a =1.6$.
The same calculations show that the approximate Eq.~\eqref{PAEH-2} for the pseudoadiabatic water warming
constitutes about $60\%$ of the exact value (this is due to neglected correlation  between $T$ and $dq^*$ when taking $\ln (T/T_{\rm b})$ out of the integral in Eq.~\eqref{PAEH-2}). As the water warming term in the pseudoadiabatic
case is an order of magnitude lower than the water lifting term, this inaccuracy is inconsequential for our estimate of $K_1$ (\ref{K1}).

\begin{table}[t]
\begin{minipage}[p]{1\textwidth}
\small
\caption{Estimates of   $H_P = \overline{z}_{\rm bc}$,  $X_P = \{T_P, z_P \}$  (\ref{Acalc}), $a$  (\ref{zP}), $\beta$ (\ref{beta}), and $K_1$ (\ref{K1})  for hydrostatic pseudoadiabatic (in bold) and reversible profiles with surface temperature $T_s$, surface pressure $p_s = 950$~hPa, and $z_{\rm b} = 1$~km. }\label{revnum}
\begin{center}
\begin{tabular}{ccccccccccccc}
    \hline \hline
     \multicolumn{1}{c}{$T_{s}$,~K}&  \multicolumn{1}{c}{$T_{\rm b}$,~K}& 
    \multicolumn{1}{c}{$z_{\rm c}$,~km}&  \multicolumn{1}{c}{$T_{\rm c}$,~K}&  \multicolumn{1}{c}{$\varepsilon_{C}$}& 
     \multicolumn{1}{c}{$z_{P}$,~km}&  \multicolumn{1}{c}{$T_{P}$,~K}&  \multicolumn{1}{c}{$\varepsilon_{P}$}&  
     \multicolumn{1}{c}{$H_{P}$,~km}&
      \multicolumn{1}{c}{$a$} & \multicolumn{1}{c}{$\beta$}&  \multicolumn{1}{c}{$K_{1}$}&  
\multicolumn{1}{c}{$\beta K_1/\varepsilon_{C}$} \\    \hline
$\mathbf{303}$ & $\mathbf{300}$ &   $\mathbf{18}$ & $\mathbf{197}$ & $\mathbf{0.34}$ & $\mathbf{11}$  & $\mathbf{275}$ & $\mathbf{0.08}$ & 
$\mathbf{6.8}$ &  $\mathbf{1.6}$ &$\mathbf{0.65}$ & $\mathbf{0.042}$ & $\mathbf{0.08}$ \\
303 & 300 &   18  & $201$ & $0.33$ & $18$ & $201$ & $0.33$ & & &$0.65$ & $0.044$ & $0.09$ \\ 
$\mathbf{303}$ & $\mathbf{300}$ &   $\mathbf{10}$ & $\mathbf{262}$ & $\mathbf{0.13}$ & $\hphantom{0}\mathbf{7}$   & $\mathbf{283}$ & $\mathbf{0.06}$ & $\mathbf{5.3}$ & $\mathbf{1.3}$ &$\mathbf{0.77}$ & $\mathbf{0.026}$ & $\mathbf{0.16}$ \\
303 & 300 &  10& $261$ & $0.13$ & $10$  & $261$ & $0.13$ & & &$0.78$ & $0.036$ & $0.22$ \\
$\mathbf{289}$ & $\mathbf{284}$ &   $\mathbf{11}$ & $\mathbf{211}$ & $\mathbf{0.26}$ & $\hphantom{100}\mathbf{5.5}$   & $\mathbf{266}$ & $\mathbf{0.06}$ &  $\mathbf{4.1}$ & $\mathbf{1.3}$ & $\mathbf{0.77}$ & $\mathbf{0.020}$ & $\mathbf{0.06}$ \\
289 & 284 &  11& $212$ & $0.25$ & $11$  & $266$ & $0.25$ & & & $0.77$ & $0.028$ & $0.09$ \\    \hline        
\end{tabular}
\end{center}
\end{minipage}
\end{table}


\vspace{0.2cm}

\subsection{Reply to the Editor and reviewers}

{\color{blue} \it

\noindent
Oct 11, 2021  

\noindent
Ref.: JAS-D-21-0172\footnote{Original submission JAS-D-21-0172 with line numbers can be found at \href{https://bioticregulation.ru/ab.php?id=mpi4}{https://bioticregulation.ru/ab.php?id=mpi4}}

\noindent
Editor Decision

\vspace{0.45cm}\noindent
Dear Dr. Makarieva,

\vspace{0.15cm}\noindent
I am now in receipt of two reviews of your manuscript "Water lifting and outflow gain of kinetic energy in tropical cyclones", and an editorial decision of Minor Revision has been reached. I apologize for the delay in decision. We had lost the help of one reviewer and therefore had to seek out an additional Associate Editor to review your manuscript. The reviews are included below. As you will see, although the decision is Minor Revision, the reviewers seek some vital clarification on key equations and the writing. Please address these as thoroughly and succinctly as possible in revision.

\vspace{0.15cm}\noindent
We invite you to submit a revised paper by 10 December 2021. If you anticipate problems meeting this deadline, please contact me as soon as possible at Rozoff.JAS@ametsoc.org.

\vspace{0.15cm}\noindent
Along with your revision, please upload a point-by-point response that satisfactorily addresses the concerns and suggestions of each reviewer. To help the reviewers and Editor assess your revisions, our journal recommends that you cut-and-paste the reviewer and Editor comments into a new document. As you would conduct a dialog with someone else, insert your responses in a different font, different font style, or different color after each comment. If you have made a change to the manuscript, please indicate where in the manuscript the change has been made.  (Indicating the line number where the change has been made would be one way, but is not the only way.)

\vspace{0.15cm}\noindent
Although our journal does not require it, you may wish to include a tracked-changes version of your manuscript. You will be able to upload this as "additional material for reviewer reference". Should you disagree with any of the proposed revisions, you will have the opportunity to explain your rationale in your response.

\vspace{0.15cm}\noindent
Please go to www.ametsoc.org/PUBSrevisions and read the AMS Guidelines for Revisions. Be sure to meet all recommendations when revising your paper to ensure the quickest processing time possible.

\vspace{0.15cm}\noindent
When you are ready to submit your revision, go to https://www.editorialmana\-ger.com/amsjas/ and log in as an Author. Click on the menu item labeled "Submissions Needing Revision" and follow the directions for submitting the file.

\vspace{0.15cm}\noindent
Thank you for submitting your manuscript to the Journal of the Atmospheric Sciences. I look forward to receiving your revision.

\vspace{0.15cm}\noindent
Sincerely,

\vspace{0.15cm}
\noindent
Dr. Christopher Rozoff

\noindent
Editor

\noindent
Journal of the Atmospheric Sciences

\vspace{0.25cm}\noindent
{\color{black}  \rm

\noindent
December 2, 2021

\noindent
Ref.: JAS-D-21-0172

\noindent
Resubmission of revised manuscript

\vspace{0.15cm}\noindent
Dear Dr. Rozoff,

\vspace{0.15cm}\noindent
Thank you for your consideration of our work and for your positive evaluation of it.
We have re-organized the manuscript following our reviewers' suggestions. We have added
a new figure, which summarizes the main equations and assumptions, as suggested
by our first reviewer.

\vspace{0.15cm}\noindent
The new framework that we outline has diverse ramifications, which cannot be all
considered in one paper. We appreciate this opportunity to present the entire framework
in a comprehensive and explicit form, such that its implications could be further developed
by anyone interested.

\vspace{0.15cm}\noindent
Sincerely,

\vspace{0.15cm}\noindent
Anastassia Makarieva 

\noindent
(on behalf of the authors)

\vspace{0.2cm}

\noindent
RESPONSE TO  THE REVIEWERS

\vspace{0.15cm}\noindent 
We sincerely appreciate the reviewers' constructive comments and suggestions. 
}

\vspace{0.15cm}\noindent
\begin{center}
Reviewer \#1: 
\end{center}

\vspace{0.15cm}\noindent
Summary: The authors start from a more general set of basic equations and tested different assumptions for deriving MPI. I think this study is beneficial for critical thinking of MPI under different assumptions and enables the quantitative comparison among the corresponding results. I can generally follow the derivations with additional effort in some places which need more clarification. My comments are mainly about these unclear/uneasy-to-follow places.

\vspace{0.35cm}
\noindent
Minor comments:

\noindent
1. L163: Eq(8) used Vb=VB' and Vc=VC', doesn't it mean an infinitely narrow bcC'B'b?

\vspace{0.15cm}\noindent 
{\color{black}  \rm We clarified in the end of section~\ref{cdt} as follows: Equation \eqref{relm} describes an infinitely narrow cycle $\rm B'bcC'B'$ with $\pt V^2/\pt z = 0$ at $\rm cC'$ and $\pt V^2/\pt r = 0$ at $\rm B'b$. It is also valid for a special case of $\rm B'bcC'B'$ being a {\it closed streamline} with $V_{\rm b} = V_{\rm B'}$ \citep{em88}.}

\vspace{0.15cm}
\noindent
2. Eq(17), Eq(25) and L709: the formula partial(Vc)/partial(r) is tricky at the first glance, I think extra explanation is needed. E.g., 'it is the result of the right hand side of Eq(16) after taking the limit'.

\vspace{0.15cm}\noindent 
{\color{black} \rm We have added a clarification after Eq.~(\ref{alg}).}

\vspace{0.15cm}
\noindent
3. Eq(21): '=' -$>$ '$\simeq$'

\vspace{0.15cm}
\noindent
{\color{black} \rm Corrected.}

\vspace{0.15cm}
\noindent
4. L255: 'in the right-hand' -$>$ 'on the right-hand'

\vspace{0.15cm}
\noindent
{\color{black}\rm Corrected everywhere.}

\vspace{0.15cm}
\noindent
5. L281-L282: Why Tw=(Tb+Tc)/2?

\vspace{0.15cm}
\noindent
{\color{black} \rm This relationship is approximate ($=$ replaced with $\simeq$ in the revised text) and valid for small $\varepsilon_C$, as explained in the Appendix.
It is intended as a possible way to understand the physical meaning of Eq.~(\ref{stc}). We have clarified as follows: "Warming the water removed at the colder isotherm with temperature $T_{\rm c}$ and returned at the warmer isotherm with $T_{\rm b}$ requires extra heat $c_l (T_{\rm b} - T_{\rm c})\Delta q^*$. 
As the moist air ascends and cools from $T_{\rm b}$ to $T_{\rm c}$, the mean temperature at which the water loses heat is, in the linear approximation, 
 $T_{\rm w} \simeq (T_{\rm b} + T_{\rm c})/2$. Accordingly, the maximum efficiency with which this heat can be converted to work is $(T_{\rm b} - T_{\rm w})/T_{\rm b} \simeq \varepsilon_C/2$."
}

\vspace{0.15cm}
\noindent
6. Eq(26) vs Eq(27): In appendix, A29, A33 and A34 -$>$ A30, A38 and A39 is not straightforward enough.

\vspace{0.15cm}
\noindent
{\color{black}\rm We have now presented these derivations in a more detailed and explicit form, see Eqs.~(\ref{A})-(\ref{zP}). This more accurate consideration produced 
a somewhat greater estimate of $z_P$ (\ref{def1}) for the pseudoadiabatic case (\ref{zP}) and brought the pseudoadiabatic and reversible values of $K_1$ closer to each other. 
Thank you for emphasizing this point in our derivations.}

\vspace{0.15cm}
\noindent
7. L330: need to mention where such point is: 'e.g., at/near RMW'

\vspace{0.15cm}
\noindent
{\color{black}\rm Clarified as suggested.}

\vspace{0.15cm}
\noindent
8. L341-L342: it's better to point out the place/assumption in Emanuel (1986) that leads to the incorrectness.

\vspace{0.15cm}
\noindent
{\color{black}\rm We clarified in a footnote as follows.
In contrast, \citet[][p.~602]{em86} incorrectly concluded that the
outflow term becomes significant if the outflow radius is very large. 
{\it Footnote:} This conclusion stemmed from
\citet{em86}{\textquoteright}s Eq.~(18), where an outflow term
proportional to a large squared radius first appeared. While deriving
this equation for a cycle with finite $\rm B'b$,
\citet[][p.~588]{em86}, on the one hand, used the conservation of
angular momentum along streamlines $\rm bc$ and $\rm B'C'$ and, on the
other hand, assumed $r_{\rm B'}$ (interpreted as
{\textquotedblleft}the radial extent of the storm near the sea
level{\textquotedblright}) to be large enough for $\pt p/\pt r$ to
vanish and, at the same time, small enough for  $r \pt p/\pt r |_{\rm
b} \gg  r \pt p/\pt r |_{\rm B'}$\citep[for details,
see][Appendix~A]{makarieva18b}. With $r \pt p/ \pt r \sim \rho v^2$,
ignoring this term at $\rm B'$ means that $V^2_{\rm b} - V^2_{\rm B'}
\sim V^2_{\rm b}$, while at the point of maximum wind this difference
is zero.}

\vspace{0.15cm}
\noindent
9. Eq(33): this is a general form for both dry and moist situations, right? If so, L361-L362 is confusing. Only sensible heat input? I think under isothermal assumption, there can still be latent heat input if the boundary layer is not saturated. 

\noindent
10. L379-L383: as the authors mentioned, (40)

\vspace{0.15cm}
\noindent
{\color{black} \rm Former Eq.~(33) (now Eq.~\ref{FV}) is indeed a general form for both dry and moist situations. It says that at the point of maximum wind, work equals dissipation. By itself, it does not say anything about heat input. However, under isothermal conditions, work $-\alpha dp$ corresponds to sensible heat input.
This is what was stated at L361-L362.

\noindent
The latent heat is indeed present (former Eq.~(40), now Eq.~\ref{sum}). Work (and dissipation at the point of maximum wind) equals total heat input times efficiency.  Total heat input is the sum of latent and sensible heat. Efficiency is less than unity. So there is a possibility for the relationship $-\alpha dp = \varepsilon (-\alpha dp + L_v dq)$  to hold. However, it cannot be generally valid. 
This is discussed in section~\ref{kk}.
}

\vspace{0.15cm}
\noindent
11. L526-L528: it is equivalent to Vc*rc-1/2f*rb$^2$ $>= 0$ given Mb=Mc. Does this constrain the selection of c point? Maybe I missed it, is this c or C' point arbitrary picked?  Can the authors comment on this c selection?

\vspace{0.15cm}
\noindent
{\color{black} \rm Generally, the selection of point $\rm c$ is completely arbitrary. Moreover, as we now clarify after Eq.~(\ref{eq2}), the obtained results
do not depend on the orientation of $\rm cC'$ (horizontal, vertical or tilted). Expression for $K_2$ \eqref{K2} is written taking into acount Eq.~\eqref{voo}, which is based on $M_{\rm b}= M_{\rm c}$. Therefore, the condition $v_{\rm c} = 0$ specifies the outflow
radius $r_{\rm c}$ according to the above relationship. It is about $r_{\rm c} \lesssim 10 r_{\rm b}$, as discussed in section~\ref{out}.
}

\vspace{0.15cm}
\noindent
12. In the appendix: are A21-A26 for reversible process? If so, please explicitly state it. It's easier for readers to get a sense quickily.

\vspace{0.15cm}
\noindent
{\color{black} \rm We clarified after Eq.~(\ref{oints}) that Eqs.~(\ref{term-1})-(\ref{dq}) are general, while before Eq.~(\ref{term-2}) we noted that Eqs.~(\ref{term-2})-(\ref{varep}) are for reversible processes.}

\vspace{0.15cm}
\noindent
13. L688: T=Tp: I think it is not very straight-forward to get this, can the authors provide some estimation about the error as for A26?

\vspace{0.15cm}
\noindent
{\color{black} \rm This error estimation (about $60\%$ but not consequential for the main conclusions) is provided after Eq.~(\ref{zP}).}

\vspace{0.15cm}
\noindent
14. This paper is not easy for readers to go though given the dense equations. I would suggest the authors emphasize a little more on the key equations and the comparisons with the previous studies. A table and a figure like table 1 and Figure 2 but for comparison of assumptions / equations will be of great help. 

\vspace{0.15cm}
\noindent
{\color{black} \rm We have added a new figure (Fig.~\ref{fig3}), where we emphasize the key assumptions and equations. 
 In the text, we emphasize that there are three distinct assumptions (Eqs.~(\ref{as1}), (\ref{as2}) and (\ref{as3})) that lead two different resulting formulae for $V_{\rm max}$. }

\vspace{0.25cm}
\begin{center}
Reviewer \#3: 
\end{center}

\vspace{0.15cm}
\noindent
Brief summary:

\vspace{0.15cm}
\noindent
This study examines the role of water lifting in tropical cyclones in a theoretical framework and finds that the water lifting's effect on reducing the storm intensity is generally small. The framework is energetically-based and axisymmetric. The relationship between the theory and Emanuel's maximum potential intensity theory is also described.

\vspace{0.15cm}
\noindent
Recommendation:

\vspace{0.15cm}
\noindent
The manuscript is well written and provides a nice theoretical framework for understanding the role of water lifting in ideal tropical cyclones. My overall recommendation is to accept, but I encourage the authors to consider the general comments below to improve the manuscript.

\vspace{0.15cm}
\noindent
General comments:

\vspace{0.15cm}
\noindent
1) The manuscript is a bit dense and difficult to follow at times. Part of this is unavoidable due to the theoretical nature of the manuscript. However, I would recommend some rewriting of the conclusions section to better bring out the main points. For example, in first paragraph of the conclusions, a main summary of the important findings is not given. Rather, first equations are given without any segue to provide context. Additionally, the conclusions section contains reviews of past work, which would be better located in the introduction or main body. Many readers of JAS will first look at the abstract and conclusions to determine interest in the rest of the paper. It is important to distill the main points in the manuscript into these sections so that the paper will be of more general use to a broader audience.

\vspace{0.15cm}
\noindent
{\color{black} \rm We have re-organized the manuscript as follows. In the Introduction, we presented a clear structure of the main findings of the paper (the last three paragraphs). In the concluding section, we have merged the material from the first three subsections ("How significant is the water lifting term?", "The estimate of Sabuwala et al. (2015)", "The outflow term") to the corresponding sections in the main body of the manuscript (now sections \ref{wlft} "Water lifting" and \ref{vmaxn} "Estimates of maximum velocity").  The flow of arguments in the Conclusions  is now facilitated by a graphical abstract of the main findings in the new Figure~\ref{fig3} suggested by the first referee.}

\vspace{0.15cm}
\noindent
2) A critical aspect of any theory is to demonstrate either through observational analysis or numerical simulations that the theory is valid. This manuscript contains no internal evaluation of the theory, although comparisons to other studies' estimates of water lifting power are given in Table 2. I understand the manuscript is already quite long, but such analysis would strengthen the manuscript. In the end, perhaps such analysis needs to be in a follow-on manuscript, but I do believe it is needed at some point. I suggest using the axisymmetric CM1 model to validate the theory in ideal conditions. For example, the water lifting term could be calculated in the model simulations and its significance could be assessed in relation to the theory.

\vspace{0.15cm}
\noindent
{\color{black} \rm In this work, we have outlined the theoretical framework as comprehensively and explicitly as possible. (We believe that it could be a certain lack of clarity that contributed to the corresponding issues having remained unsettled.) This has indeed made the manuscript quite long.  However, as we state in the revised Conclusions: "The developed analytical framework can be used to evaluate corresponding magnitudes for different  scenarios in numerical models." This can be done in a follow-on study.}

\vspace{0.15cm}
\noindent
Specific minor comments:

\vspace{0.15cm}
\noindent
L449: typo "overall"
}
{\color{black} Corrected.}

\bibliographystyle{copernicus}

\nopagebreak

\end{document}